\PassOptionsToPackage{unicode}{hyperref}
\PassOptionsToPackage{hyphens}{url}
\PassOptionsToPackage{dvipsnames,svgnames,x11names}{xcolor}
\documentclass[
  12pt]{article}

\usepackage{amsmath,amssymb}
\usepackage{algorithm}
\usepackage{algpseudocode}
\usepackage{caption, booktabs, xcolor}
\usepackage{iftex}
\ifPDFTeX
  \usepackage[T1]{fontenc}
  \usepackage[utf8]{inputenc}
  \usepackage{textcomp} 
\else 
  \usepackage{unicode-math}
  \defaultfontfeatures{Scale=MatchLowercase}
  \defaultfontfeatures[\rmfamily]{Ligatures=TeX,Scale=1}
\fi
\usepackage{lmodern}
\ifPDFTeX\else  
\fi
\IfFileExists{upquote.sty}{\usepackage{upquote}}{}
\IfFileExists{microtype.sty}{
  \usepackage[]{microtype}
  \UseMicrotypeSet[protrusion]{basicmath} 
}{}
\makeatletter
\@ifundefined{KOMAClassName}{
  \IfFileExists{parskip.sty}{%
    \usepackage{parskip}
  }{
    \setlength{\parindent}{0pt}
    \setlength{\parskip}{6pt plus 2pt minus 1pt}}
}{
  \KOMAoptions{parskip=half}}
\makeatother
\usepackage{xcolor}
\makeatletter
\ifx\paragraph\undefined\else
  \let\oldparagraph\paragraph
  \renewcommand{\paragraph}{
    \@ifstar
      \xxxParagraphStar
      \xxxParagraphNoStar
  }
  \newcommand{\xxxParagraphStar}[1]{\oldparagraph*{#1}\mbox{}}
  \newcommand{\xxxParagraphNoStar}[1]{\oldparagraph{#1}\mbox{}}
\fi
\ifx\subparagraph\undefined\else
  \let\oldsubparagraph\subparagraph
  \renewcommand{\subparagraph}{
    \@ifstar
      \xxxSubParagraphStar
      \xxxSubParagraphNoStar
  }
  \newcommand{\xxxSubParagraphStar}[1]{\oldsubparagraph*{#1}\mbox{}}
  \newcommand{\xxxSubParagraphNoStar}[1]{\oldsubparagraph{#1}\mbox{}}
\fi
\makeatother

\usepackage{longtable,booktabs,array}
\usepackage{calc} 
\usepackage{etoolbox}
\makeatletter
\patchcmd\longtable{\par}{\if@noskipsec\mbox{}\fi\par}{}{}
\makeatother
\IfFileExists{footnotehyper.sty}{\usepackage{footnotehyper}}{\usepackage{footnote}}
\makesavenoteenv{longtable}
\usepackage{graphicx}
\makeatletter
\def\maxwidth{\ifdim\Gin@nat@width>\linewidth\linewidth\else\Gin@nat@width\fi}
\def\maxheight{\ifdim\Gin@nat@height>\textheight\textheight\else\Gin@nat@height\fi}
\makeatother
\setkeys{Gin}{width=\maxwidth,height=\maxheight,keepaspectratio}
\makeatletter
\def\fps@figure{htbp}
\makeatother

\makeatletter
\@ifpackageloaded{caption}{}{\usepackage{caption}}
\AtBeginDocument{%
\ifdefined\contentsname
  \renewcommand*\contentsname{Table of contents}
\else
  \newcommand\contentsname{Table of contents}
\fi
\ifdefined\listfigurename
  \renewcommand*\listfigurename{List of Figures}
\else
  \newcommand\listfigurename{List of Figures}
\fi
\ifdefined\listtablename
  \renewcommand*\listtablename{List of Tables}
\else
  \newcommand\listtablename{List of Tables}
\fi
\ifdefined\figurename
  \renewcommand*\figurename{Figure}
\else
  \newcommand\figurename{Figure}
\fi
\ifdefined\tablename
  \renewcommand*\tablename{Table}
\else
  \newcommand\tablename{Table}
\fi
}
\@ifpackageloaded{float}{}{\usepackage{float}}
\floatstyle{ruled}
\@ifundefined{c@chapter}{\newfloat{codelisting}{h}{lop}}{\newfloat{codelisting}{h}{lop}[chapter]}
\floatname{codelisting}{Listing}

\makeatother
\makeatletter
\@ifpackageloaded{caption}{}{\usepackage{caption}}
\@ifpackageloaded{subcaption}{}{\usepackage{subcaption}}
\makeatother

\ifLuaTeX
  \usepackage{selnolig}  
\fi
\usepackage[]{natbib}
\usepackage{bookmark}

\IfFileExists{xurl.sty}{\usepackage{xurl}}{} 
\hypersetup{
  pdftitle={Title},
  pdfauthor={Author 1; Author 2},
  pdfkeywords={3 to 6 keywords, that do not appear in the title},
  colorlinks=true,
  linkcolor={blue},
  filecolor={Maroon},
  citecolor={Blue},
  urlcolor={Blue},
  pdfcreator={LaTeX via pandoc}}

\newcommand{\anon}{1}

\begin{document}

\def\spacingset#1{\renewcommand{\baselinestretch}%
{#1}\small\normalsize} \spacingset{1}


\if1\anon
{
  \title{\bf Component-weighted sparse group LASSO for finite mixture regression  models}
  \author{Vinay Joshy 1\\
    Department of Mathematics and Statistics, University of Guelph\\
    and \\
    Zeny Feng 2\thanks{Corresponding author. Email: zfeng@uoguelph.ca}\\
    Department of Mathematics and Statistics, University of Guelph\\
    and \\
    Grace Stelter 3 \\
    Department of Mathematics and Statistics, University of Guelph\\
    and \\
    Lorna E. Deeth 4 \\
    Department of Mathematics and Statistics, University of Guelph\\
    and \\
    Alysha Cooper 5 \\
    Department of Mathematics and Statistics, University of Guelph}
  \maketitle
} \fi

\if0\anon
{
  \bigskip
  \bigskip
  \bigskip
  \begin{center}
    {\LARGE\bf Component-weighted sparse group LASSO for finite mixture regression  models}
\end{center}
  \medskip
} \fi

\bigskip
\begin{abstract}
When analyzing heterogeneous data with latent subpopulations, finite mixture regression (FMR) models are effective as they allow for variations in regression coefficients across mixture components. Variable selection is important at two levels: (1) at the group level by removing completely irrelevant covariates across all subpopulations and (2) at the individual level by removing irrelevant covariates within each subpopulation. However, existing variable selection methods for FMR focus solely on individual-level regression coefficient selection, providing no mechanism for group-level variable elimination. We introduce a component-weighted sparse group LASSO regularization method that performs efficient variable selection at both levels simultaneously in finite mixture regression models belonging to the exponential family. We develop a novel optimization procedure through a Majorization-Minimization algorithm. Simulation studies under Gaussian and Poisson mixture regression settings demonstrate reliable recovery of the true sparsity structure at both levels. The method is illustrated by analyzing two real datasets.
\end{abstract}

\noindent%
\vfill

\newpage
\spacingset{1.8} 

\section{Introduction \label{sec:intro}}
Finite mixture models provide a flexible probabilistic framework for representing data arising from heterogeneous populations, where observations are assumed to originate from one of $G$ latent subpopulations, each governed by its own distribution \citep{mclachlanFiniteMixtureModels2000, titteringtonStatisticalAnalysisFinite1985a}. The statistical properties of these models, including identifiability and consistency of maximum likelihood estimators, have been studied extensively \citep{rednerMixtureDensitiesMaximum1984, lerouxConsistentEstimationMixing1992}. When observations depend on covariates, the finite mixture framework extends naturally to finite mixture regression (FMR) models, in which each component has its own regression relationship between the response and covariates. 

\citet{wedelMixtureLikelihoodApproach1995} established a general framework for finite mixtures of generalized linear models (GLMs) where an Expectation-Maximization (EM) algorithm \citep{dempsterMaximumLikelihoodIncomplete1977} provides the standard approach for maximum likelihood estimation of its parameters. Finite mixture and FMR models have found application across a wide range of disciplines. In biology and ecology, they have been used to identify morphologically or behaviorally distinct subpopulations within a species \citep{mclachlanFiniteMixtureModels2000}. In the social sciences and market research, FMR models provide a model-based approach to market segmentation, identifying consumer segments with distinct preference structures \citep{wedelMarketSegmentation2000}.

As the number of covariates $p$ grows, a fundamental challenge in FMR modeling is variable selection: not all covariates are relevant to the response in every component, and including irrelevant covariates inflates variance and reduces interpretability. Classical information criteria such as Akaike's information criterion~\citep{akaikeInformationTheoryExtension1998} and Bayesian information criterion (BIC) ~\citep{schwarzEstimatingDimensionModel1978} become computationally intractable as $p$ and $G$ increase, since they require fitting all possible submodels. Penalized likelihood methods address this directly by regularizing the parameter estimates during estimation. \citet{khalili2007} introduced a class of penalties for FMR models that accounts for the mixture structure, proposing MIXLASSO and MIXSCAD estimators and establishing consistency of the penalized estimator for variable selection. \citet{stadler1penalizationMixtureRegression} further studied $\ell_1$-penalized estimation in Gaussian mixture regression. These contributions represent significant advances; however, both frameworks operate at the level of individual coefficients and do not explicitly enforce that a covariate is entirely absent from the model i.e., shrinks all of its regression coefficients to zero across all components simultaneously.

This motivates a two-level variable selection problem: identifying covariates that are irrelevant in specific components (individual-level sparsity), while also identifying covariates that contribute nothing across all components (group-level sparsity). The two goals are distinct, and achieving both simultaneously requires a structured penalty that operates at both levels.

\citet{simonSparseGroupLasso2013} introduced the Sparse Group LASSO (SGL) for linear regression, which combines an $\ell_1$ (LASSO) penalty~\citep{tibshirani1996} for individual-level sparsity with a group LASSO penalty~\citep{yuanModelSelectionEstimation2006a} for group-level sparsity, enabling simultaneous selection at both levels. In the standard SGL setting, groups are defined by the analyst as meaningful subsets of covariates. The SGL has demonstrated strong empirical performance in structured variable selection problems, and its theoretical properties in the non-mixture setting are well understood. 

In this paper, we combine the ideas of \citet{khalili2007} and \citet{simonSparseGroupLasso2013} to propose a component-weighted Sparse Group LASSO penalty, which we will refer to as $\pi$-SGL, for variable selection in FMR models. Throughout, we use \emph{component} to refer to a latent subpopulation in the
mixture, and reserve \emph{group} for a block of regression coefficients that the
penalty treats as a single unit, following the group LASSO convention. In the mixture regression setting, the natural grouping structure is defined column-wise: each group collects the coefficient for a given covariate across all $G$ components, so that the group penalty can shrink a covariate globally when it is irrelevant in every subpopulation. The mixing proportions, $\pi_g$, serve as component-specific weights, reflecting the relative sample size of each latent subpopulation and hence its influence on the penalty.

The $\pi$-SGL penalized likelihood is non-smooth, and the standard EM algorithm does not directly accommodate non-smooth penalties without modification. To address this, we develop an Majorization-Minimization (MM) algorithm \citep{hunterTutorialMMAlgorithms2004a} that constructs a tractable surrogate majorizing the penalized objective at each iteration, reducing each MM step to an iteratively reweighted penalized least squares procedure with closed-form solutions.

We evaluate the proposed method through simulation studies under both Gaussian and Poisson finite mixture regression settings, assessing variable selection performance at both the group and individual levels across a range of sample sizes, penalty configurations, and signal strengths. We further demonstrate the method on two real data applications. The remainder of the paper is organized as follows. Section 2 introduces finite mixture of GLMs and the $\pi$-SGL penalty, and derives the MM algorithm. Section 3 describes the simulation studies and reports variable selection performance. Section 4 presents the real data applications. Section 5 concludes with a discussion of findings, limitations, and future directions.

\section{Methods}\label{sec:methods}

Let $\mathbf{y} = (y_1, \ldots, y_n)^\top$ denote the response vector and $\mathbf{X}$ be an $n \times (p+1)$ matrix of an intercept and $p$ covariates. For data arising from a mixture of $G$ distributions, the response $y_i$ has density:
\begin{equation*}
    f(y_i; \mathbf{x}_i, \Psi) = \sum_{g=1}^{G} \pi_g f_g(y_i; \mathbf{x}_i, \psi_g),
\end{equation*}

where $\pi_g > 0$ are mixing proportions satisfying $\sum_{g=1}^{G} \pi_g = 1$, $f_g(y_i; \mathbf{x}_i, \psi_g)$ is the $g$-th component density from a parametric family, and $\psi_g$ collects the component-specific parameters. Each component belongs to the exponential family with density:
\begin{equation*}
    f_g(y_i; \theta_{ig}, \phi) = \exp\left\{ \frac{y_i \theta_{ig} - b(\theta_{ig})}{a(\phi)} + c(y_i, \phi) \right\},
\end{equation*}

where $\phi$ is the dispersion parameter and $a(\cdot)$, $b(\cdot)$, $c(\cdot)$ are known functions. Under the canonical link $h(\cdot)$, the natural parameter $\theta_{ig}$ is equal to the linear predictor:
\begin{equation*}
    \theta_{ig} = \eta_{ig} = h(\mu_{ig}) = \mathbf{x}_i^\top \boldsymbol{\beta}_g,
\end{equation*}

where $\mu_{ig} = \mathbb{E}[y_i \mid \theta_{ig}] = b'(\theta_{ig})$. The incomplete data log-likelihood is:
\begin{equation*}
    \ell(\Psi) = \sum_{i=1}^{n} \log \sum_{g=1}^{G} \pi_g f_g(y_i; \mathbf{x}_i, \psi_g),
\end{equation*}

where $\Psi = \{\pi_g, \psi_g\}_{g=1}^G$ denotes the full parameter vector with $\psi_g=(\beta_g, \phi_g)$. Unlike standard regularization methods, in finite mixture models we aim to identify covariates that are irrelevant either across all mixture components or specific to certain components while remaining relevant in others. We propose the $\pi$-dependent Sparse Group LASSO ($\pi$-SGL) penalty:
\begin{equation*}
    J(\boldsymbol{\beta}) = (1-\alpha)\sum_{j=1}^{p}\sqrt{G}\sqrt{\sum_{g=1}^{G}\pi_g^2\beta_{jg}^2} + \alpha\sum_{j=1}^{p}\sum_{g=1}^{G}\pi_g|\beta_{jg}|, \quad 0 \leq \alpha \leq 1
\end{equation*}

where the first term encourages group-wise regularization of the $j$-th covariate across all $G$ components, and the second term allows for element-wise sparsity for specific components. The mixing proportions, $\pi_g$, serve as weights, scaling each term by the sample size in their respective subpopulation in the mixture. A subpopulation with a smaller size will have less harsh penalties applied. The tuning parameter $\alpha$ governs the trade-off between the two levels of sparsity: when $\alpha = 1$, $J(\boldsymbol{\beta})$ reduces to the $\pi$-weighted LASSO, and when $\alpha = 0$, it reduces to the $\pi$-weighted group LASSO.

The objective function to be minimized is:
\begin{equation*}
    Q(\Psi) = -\ell(\Psi) + \lambda J(\boldsymbol{\beta}), \quad \lambda \in [0, \lambda_{\max}]
\end{equation*}

where $\lambda \geq 0$ controls the overall strength of regularization and $\lambda_{\max}$ is the smallest value of $\lambda$ that shrinks all non-intercept coefficients to zero.

The objective function $Q(\Psi)$ is non-convex and non-differentiable due to the mixture structure and regularization terms. We employ the MM algorithm, which iteratively minimizes a surrogate function $h(\Psi|\Psi^{(t)})$ that majorizes $Q(\Psi)$ at the current iterate $\Psi^{(t)}$, satisfying:
\begin{equation}
    Q(\Psi) \leq h(\Psi|\Psi^{(t)}) \quad \forall\, \Psi, \qquad Q(\Psi^{(t)}) = h(\Psi^{(t)}|\Psi^{(t)}).
\end{equation}

At the majorization step, the surrogate is constructed as the sum of two separate majorizing functions, one for the negative log-likelihood $-\ell(\Psi)$ and the other for the penalty $\lambda J(\boldsymbol{\beta})$. Applying Jensen's inequality, we obtain the surrogate of $-\ell(\Psi)$ as (see Appendix~\ref{app:jen_inq}):
\begin{equation*}
    h_\ell(\Psi|\Psi^{(t)}) = -\sum_{i=1}^{n}\sum_{g=1}^{G} z_{ig}^{(t)} \left(\log \pi_g + \frac{y_i \theta_{ig} - b(\theta_{ig})}{a(\phi)}\right) + C_\ell^{(t)}, 
\end{equation*}

where $C_\ell^{(t)}$ is a constant with respect to $\Psi$, and $z_{ig}^{(t)}$ turns out to be equivalent to the posterior probability that observation $i$ belongs to component $g$ given $\psi_g^{(t)}$:
\begin{equation*}
    z_{ig}^{(t)} = \frac{\pi_g^{(t)} f_g(y_i; \mathbf{x}_i, \psi_g^{(t)})}{\sum_{g'=1}^{G} \pi_{g'}^{(t)} f_{g'}(y_i; \mathbf{x}_i, \psi_{g'}^{(t)})}.
\end{equation*}

Employing the dominating hyperplane inequality as in \citet{cooperDominatingHyperplaneRegularization2025} for $J(\boldsymbol{\beta})$, we obtain (see Appendix~\ref{app:dhi}):
\begin{equation}\label{eqn:surr_pen}
    h_J(\boldsymbol{\beta}|\boldsymbol{\beta}^{(t)}) = \lambda\sum_{j=1}^{p}\sum_{g=1}^{G} v_{jg}^{(t)} \pi_g^2 \beta_{jg}^2 + C_J^{(t)}
\end{equation}

where $C_J^{(t)}$ is a constant with respect to all coefficients $\boldsymbol{\beta}$, and:
\begin{equation}\label{eqn:v_fxn}
    v_{jg}^{(t)} = \frac{(1-\alpha)\sqrt{G}}{2\sqrt{\sum_{g'=1}^{G}(\pi_{g'}^{(t)}\beta_{jg'}^{(t)})^2} } + \frac{\alpha}{2\sqrt{(\pi_g^{(t)}\beta_{jg}^{(t)})^2}}.
\end{equation}

The two denominators in Eqn~\ref{eqn:v_fxn} vanish as the corresponding coefficients shrink to zero, the first when the entire $j$-th covariate collapses as $\boldsymbol{\beta}_j \rightarrow \boldsymbol{0}$ and the second when $\beta_{jg} \rightarrow 0$, causing $v_{jg}^{(t)}$ to be undefined at the sparse solutions that the penalty targets. To keep the surrogate well-defined throughout iterations, we add a small constant $\varepsilon_v > 0$ to each denominator, following the perturbed majorization device of \citet{hunterVariableSelectionUsing2005}. As they note, the perturbed weights no longer majorize $Q(\Psi)$ exactly; instead they majorize a perturbed objective $Q_{\varepsilon_v}(\Psi)$ for which the descent property holds, and $|Q_{\varepsilon_v}(\Psi) - Q(\Psi)| \to 0$ uniformly on compact subsets of the parameter space as $\varepsilon_v \to 0$ (their Proposition 3.2). The perturbation acts only on the surrogate weights, not on the objective being optimized. We arbitrarily fix $\varepsilon_v = 10^{-10}$ in the numerical examples of Sections~\ref{sec:sim} and \ref{sec:data}.

Combining both surrogates, we obtain $h(\Psi|\Psi^{(t)}) = h_\ell(\Psi|\Psi^{(t)}) + h_J(\boldsymbol{\beta}|\boldsymbol{\beta}^{(t)})$, which will be minimized at the minimization step of each iteration $t$. The solution of $\pi_g$ is obtained by minimizing $h_\ell$ subject to the simplex constraint:
\begin{equation*}
    \pi_g^{(t+1)} = \frac{1}{n}\sum_{i=1}^{n} z_{ig}^{(t)}.
\end{equation*}

Since $\mu_{ig} = b'(\mathbf{x}_i^\top \boldsymbol{\beta}_g)$ is nonlinear in $\boldsymbol{\beta}_g$, a closed-form solution does not generally exist. We apply a second majorization via a first-order Taylor expansion of $\mu_{ig}$ around $\boldsymbol{\beta}_g^{(t)}$ (see Appendix~\ref{app:irls}), defining working weights and a working response respectively:
\begin{equation*}
    w_{ig}^{(t)} = z_{ig}^{(t)} \cdot \mathsf{v}(\mu_{ig}^{(t)}), \qquad \tilde{y}_{ig}^{(t)} = \mathbf{x}_i^\top \boldsymbol{\beta}_g^{(t)} + \frac{y_i - \mu_{ig}^{(t)}}{\mathsf{v}(\mu_{ig}^{(t)})},
\end{equation*}
where $\mathsf{v}(\mu_{ig}^{(t)})$ is the variance function evaluated at the current mean of the response, $\mu_{ig}^{(t)}$. The solution of $\boldsymbol{\beta}_g$ takes the form of:
\begin{equation*}
    \boldsymbol{\beta}_g^{(t+1)} = \left(\mathbf{X}^\top \mathbf{W}_g^{(t)} \mathbf{X} + 2\lambda a(\phi^{(t)}) \mathbf{V}_g^{(t)} \pi_g^{2(t+1)}\right)^{-1} \mathbf{X}^\top \mathbf{W}_g^{(t)} \tilde{\mathbf{y}}_g^{(t)},
\end{equation*}

where $\mathbf{W}_g^{(t)} = \mathrm{diag}(w_{1g}^{(t)}, \ldots, w_{ng}^{(t)})$ and $\mathbf{V}_g^{(t)} = \mathrm{diag}(0, v_{1g}^{(t)}, \ldots, v_{pg}^{(t)})$ with a zero in the intercept position to leave it unpenalized. The algorithm iterates until $|Q(\Psi^{(t+1)}) - Q(\Psi^{(t)})| \leq \epsilon$. All analyses in this paper used a convergence tolerance of $\epsilon =10^{-6}$. Following convergence, a regression coefficient $\hat{\beta}_{jg}$ is reported to be zero if $|\hat{\beta}_{jg}| \leq 10^{-10}$, effectively treating it as irrelevant to component $g$. Our developed MM algorithm is outlined in Algorithm~\ref{alg:mm_fmglm}.

\begin{algorithm}[H]
\caption{MM Algorithm for $\pi$-SGL Finite Mixture of GLMs}
\label{alg:mm_fmglm}
\begin{algorithmic}[1]
\Require Initial values of $\Psi^{(0)}=( \boldsymbol{\pi}^{(0)}, \boldsymbol{\beta}^{(0)}, \phi^{(0)}) $, tuning parameters $\lambda$ and $\alpha$, tolerance $\epsilon$, set $t = 0$.
\Repeat
    \For{$g = 1$ to $G$}
        \State Compute posterior probabilities:
        $$z_{ig}^{(t)} = \frac{\pi_g^{(t)} f_g(y_i; \mathbf{x}_i, \psi_g^{(t)})}{\sum_{g'=1}^{G} \pi_{g'}^{(t)} f_{g'}(y_i; \mathbf{x}_i, \psi_{g'}^{(t)})}$$
        \State Compute MM penalty weights $v_{jg}^{(t)}$ for $j = 1, \ldots, p$
        \State Update mixing proportions:
        $$\pi_g^{(t + 1)} = \frac{1}{n}\sum_{i=1}^{n} z_{ig}^{(t)}$$
        \State Compute working weights $w_{ig}^{(t)} = z_{ig}^{(t)} \cdot \mathsf{v}(\mu_{ig}^{(t)})$, working responses $\tilde{y}_{ig}^{(t)}$ and 
        \Statex \hspace{\algorithmicindent} \quad $\phi^{(t)}$ can be obtained depending on the distribution.
        \State Construct $\mathbf{W}_g^{(t)}$ and $\mathbf{V}_g^{(t)}$
        \State Update regression coefficients:
        $$\boldsymbol{\beta}_g^{(t + 1)} \leftarrow \left(\mathbf{X}^\top \mathbf{W}_g^{(t)} \mathbf{X} + 2\lambda a(\phi^{(t)}) \mathbf{V}_g^{(t)} \pi_g^{2(t + 1)}\right)^{-1} \mathbf{X}^\top \mathbf{W}_g^{(t)} \tilde{\mathbf{y}}_g^{(t)}$$
    \EndFor
    \State $t \leftarrow t + 1$
\Until{$|Q(\Psi^{(t)}) - Q(\Psi^{(t-1)})| \leq \epsilon$}
\State \Return $\hat{\Psi} = \Psi^{(t)}$
\end{algorithmic}
\end{algorithm}

\subsection{Special case in Gaussian mixtures}
When each component density $f_g$ is Gaussian, the mixture model reduces to the Gaussian mixture regression (GMR) model:
\begin{equation*}
    f(y_i; \mathbf{x}_i, \Psi) = \sum_{g=1}^{G} \pi_g \frac{1}{\sqrt{2\pi\sigma_g^2}} \exp\left\{-\frac{(y_i - \mathbf{x}_i^\top \boldsymbol{\beta}_g)^2}{2\sigma_g^2}\right\},
\end{equation*}
where $\psi_g = (\boldsymbol{\beta}_g, \sigma_g^2)$ and the full parameter vector is $\Psi = \{\pi_g, \boldsymbol{\beta}_g, \sigma_g^2\}_{g=1}^G$. Under the identity canonical link, $\mu_{ig} = \mathbf{x}_i^\top \boldsymbol{\beta}_g$, the variance function is $\mathsf{v}(\mu) = 1$, and the dispersion parameter $\phi = \sigma_g^2$ such that $a(\phi) = \phi = \sigma_g^2$.

A well-known issue in Gaussian mixture models is the unboundedness of the log-likelihood when $\sigma_g^2 \rightarrow 0$ which produces spurious artificially large maxima of the log-likelihood. This issue is pronounced in variable selection when regularization is used. To circumvent this, we follow \citet{chenINFERENCENORMALMIXTURES2008} by augmenting the log-likelihood with a penalty on the component variances as:
\begin{equation*}
    \ell_{\text{pen}}(\Psi) = \sum_{i=1}^{n} \log \left[\sum_{g=1}^{G} \pi_g f_g(y_i; \mathbf{x}_i^\top \boldsymbol{\beta}_g, \sigma_g^2)\right] + p_n(\boldsymbol{\sigma}^2),
\end{equation*}
where $p_n(\boldsymbol{\sigma}^2) = -\frac{1}{n}\sum_{g=1}^{G}\left\{S_y \sigma_g^{-2} + \log(\sigma_g^2)\right\}$ and $S_y$ is the variance of 25\% to 75\% quantile sample of $\mathbf{y}$. The penalized objective function then becomes:
\begin{equation*}
    Q(\Psi) = -\ell_{\text{pen}}(\Psi) + \lambda J(\boldsymbol{\beta}).
\end{equation*}

Since the Gaussian identity link yields $\mathsf{v}(\mu) = 1$, the working weights simplify to $w_{ig}^{(t)} = z_{ig}^{(t)}$ and the IRLS step required in the general GLM case is unnecessary, leading to faster convergence relative to the general algorithm. The $\boldsymbol{\beta}_g$ solution reduces to the solution of a penalized weighted ridge least squares regression in the form of:
\begin{equation*}
    \boldsymbol{\beta}_g^{(t+1)} = \left(\mathbf{X}^\top \mathbf{Z}_g^{(t)} \mathbf{X} + 2\lambda \sigma_g^{2(t)} \mathbf{V}_g^{(t)} \pi_g^{2(t+1)}\right)^{-1} \mathbf{X}^\top \mathbf{Z}_g^{(t)} \mathbf{y},
\end{equation*}
where $\mathbf{Z}_g^{(t)} = \mathrm{diag}(z_{1g}^{(t)}, \ldots, z_{ng}^{(t)})$. The solution of $\sigma_g^2$ has the closed form as:
\begin{equation*}
    \sigma_g^{2(t+1)} = \frac{(2S_y/n) + \sum_{i=1}^{n} z_{ig}^{(t)}(y_i - \mathbf{x}_i^\top \boldsymbol{\beta}_g^{(t+1)})^2}{\sum_{i=1}^{n} z_{ig}^{(t)} + (2/n)}.
\end{equation*}

Another way to prevent variance degeneracy is to constrain $\sigma^2$ to be common across all components. The variance penalty $p_n(\boldsymbol{\sigma}^2)$ is no longer needed and the objective reverts to:
\begin{equation*}
    Q(\Psi) = -\ell(\Psi) + \lambda J(\boldsymbol{\beta}), 
\end{equation*}
and the common variance solution becomes:
\begin{equation*}
    \sigma^{2(t+1)} = \frac{\sum_{g=1}^{G}\sum_{i=1}^{n} z_{ig}^{(t)}(y_i - \mathbf{x}_i^\top \boldsymbol{\beta}_g^{(t+1)})^2}{\sum_{g=1}^{G}\sum_{i=1}^{n} z_{ig}^{(t)}}
\end{equation*}

\subsection{Tuning and model selection}
The $\pi$-SGL penalty involves two tuning parameters: $\lambda$, which controls the overall strength of regularization, and $\alpha$, which governs the balance between group-wise and element-wise sparsity. In addition, the number of mixture components $G$ must be selected. We perform joint selection of $G$, $\lambda$, and $\alpha$ using BIC:
\begin{equation*}
    \text{BIC}(G, \lambda, \alpha) = -2\ell(\hat{\Psi}) + \log(n) \cdot k , 
\end{equation*}

where $\ell(\hat{\Psi})$ is the log-likelihood at $\hat{\Psi} $ and $k$ is the number of nonzero coefficients, intercepts, dispersion parameters and $G-1$ free mixing proportions in the fitted model. The model minimizing BIC over a candidate set of values for $G$, $\lambda$, and $\alpha$ is selected as the final model. The use of BIC for selecting $G$ is supported by \citet{keribin2000consistent}, who showed that under certain regularity conditions, BIC consistently estimates the order of a finite mixture model. 

For a fixed $G$ and $\alpha$, the solution path is computed over a decreasing grid of $\lambda$ values on the interval $[\lambda_{\max} \cdot \delta, \lambda_{\max}]$, where $0 < \delta < 1$. Adapting the approach of \citet{friedmanRegularizationPathsGeneralized2010} to the finite mixture GLM setting, $\lambda_{\max}$ is given by:
\begin{equation*}
    \lambda_{\max} = \frac{\sqrt{\sum_{g'=1}^{G}(\pi_{g'}^{(0)})^2}}
    {\sqrt{G}\,(\pi_{\min}^{(0)})^2} \max_{j,g} 
    \left|\mathbf{X}_j^\top \mathbf{Z}_g^{(0)}\left(\mathbf{y} - 
    \boldsymbol{\mu}_g^{(0)}\right)\right|,
\end{equation*}

where $\mathbf{X}_j$ is the $j$-th column of $\mathbf{X}$, $\boldsymbol{\mu}_g^{(0)}$ is the vector of estimated component means, $\mathbf{Z}_g^{(0)}$ is the matrix of posterior weights, and $\pi_{\text{min}}^{(0)}$ is the smallest estimated mixing proportion. $\boldsymbol{\mu}_g^{(0)}, \mathbf{Z}_g^{(0)}, \pi_{\text{min}}^{(0)}$ are solutions of $\boldsymbol{\mu}_g, \mathbf{Z}_g, \min(\pi_1, \dots, \pi_G )$ respectively when no covariates are included, i.e., the maximum likelihood solution of a finite mixture without regression.

\section{Simulation study}\label{sec:sim}
Two simulation studies were conducted to evaluate the performance of the proposed regularization method. For these simulation studies, a grid of values for the tuning parameter, $\lambda$, was defined over the interval [$0.001\lambda_{\max}, \lambda_{\max}$], where $\lambda_{\max}$ was determined according to Appendix~\ref{app:lambda}. All simulations were conducted in R version 4.2.1 \citep{Rcore2022}.

\subsection{Evaluation metrics}
Variable selection performance are evaluated through precision and recall metrics at both the group and within-group levels. For a given covariate $j$ in component $g$, $\beta_{jg}$ is classified as a true positive (TP) if the true coefficient is nonzero and the estimated coefficient is nonzero, a false positive (FP) if the true coefficient is zero but the estimated coefficient is nonzero, and a false negative (FN) if the true coefficient is nonzero but the estimated coefficient is shrunk to zero. A coefficient is considered nonzero if its absolute value exceeds $10^{-10}$. At the group level, a covariate $j$ is considered relevant if at least one component has a nonzero coefficient, i.e., $\underset{g}{\max}|\beta_{jg}| > 10^{-10}$. A group-level true positive occurs when a relevant covariate is retained by the model in at least one component, a false positive when an irrelevant covariate is retained in any component, and a false negative when a relevant covariate is entirely zeroed out across all components. The group-level metrics are therefore component-agnostic by construction, assessing only whether a covariate is retained somewhere in the model; whether it is retained in the correct component is assessed by the within-group metrics, under which a covariate retained in a component where it is truly zero is charged as a within-group false positive.

For each level, we report precision, recall, and their harmonic mean (F1-score):
\begin{equation*}
\text{Precision} = \frac{TP}{TP + FP}, \quad \text{Recall} = \frac{TP}{TP + FN}, \quad F1 = \frac{2 \cdot \text{Precision} \cdot \text{Recall}}{\text{Precision} + \text{Recall}}.    
\end{equation*}
Precision measures the proportion of selected coefficients (or covariates, at the group level) that are truly nonzero, while recall measures the proportion of truly nonzero coefficients (or covariates) that are correctly identified by the model. To account for label switching, estimated components were matched to true components sequentially by selecting the pair $(g, \hat{g})$ with minimum $L_1$ distance between assignment vectors $\mathbf{z}_g$ and $\hat{\mathbf{z}}_{\hat{g}}$, with matched pairs removed from consideration at each step until all components are assigned. We additionally report direction accuracy, measured as the proportion of correctly signed estimates among coefficients identified as nonzero in both the true and estimated models. Finally, we report the proportion of replications selecting the correct order, $\hat{G} = G$, via BIC.

\subsection{Gaussian mixture regression}\label{sec:gmr_sim}
Data were generated from a $G$-component Gaussian mixture regression (GMR) model with $p$ covariates. The design matrix $\mathbf{X}$ was drawn from a multivariate normal distribution with a correlation structure, $\text{Corr}(X_j, X_k) = \rho^{|j-k|}$, with $\rho = 0.2$. Component memberships were drawn from a multinomial distribution with mixing proportions $\boldsymbol{\pi}$, and responses generated as $y_i \sim \mathcal{N}(\mathbf{x}_i^\top \boldsymbol{\beta}_g, \sigma_g^2)$ for the assigned component $g$. Mixing proportions were either set equal across components, i.e., $\pi_g = 1/G$ for all $g$, or generated as $\pi_g \propto \sqrt{c_g}$ where $c_1 > c_2 > \cdots > c_G$ are $G$ equally spaced values on $[0.1, 1.0]$, normalized to sum to one, yielding decreasing proportions across components. Component standard deviations were either set equal, $\sigma_g = 0.5$ for all $g$, or generated as $\sigma_g = \sqrt{s_g}$ where $s_1 > s_2 > \cdots > s_G$ are $G$ equally spaced values on $[0.1, 1.0]$, also yielding decreasing values across components.

\begin{table}[H]
    \centering
    \caption{Gaussian mixture regression simulation design. Each scenario was replicated 100 times with fixed true parameters across replications}
    \begin{tabular}{ll}
        \hline
         Simulation Parameter & Values\\
         \hline
         Number of components ($G$) & 3, 4 \\
         Number of covariates ($p$) & 10, 25\\
         Sample size ($n$) & 300, 500 \\
         Mixing proportions ($\boldsymbol{\pi}$) & Equal, Unequal\\
         Component variances ($\boldsymbol{\sigma}^2$)& Equal, Unequal\\
         Relevant covariate proportion ($\delta_p$) & 0.3, 0.5\\
         Relevant component proportion ($\delta_w$) & 0.3, 0.5\\
         \hline
    \end{tabular}
    \label{tab:gmr_design}
\end{table}

The true coefficient matrix was constructed as follows. Of the $p$ covariates, $\lceil p \cdot \delta_p \rceil$ were designated as relevant. Among these, at least one covariate was assigned to be nonzero across all $G$ components, while the remaining relevant covariates were nonzero in $\lceil G \cdot \delta_w \rceil$ components each, selected at random. All nonzero coefficients were drawn uniformly from $[0.3, 1.0]$ with random sign. Once coefficients are generated (for each simulation scenario), they are fixed for all replicates. Irrelevant covariates had zero coefficients in all components. Component intercepts were spaced evenly to ensure separation between components and were within $[-3, 3]$.

\subsubsection{Results}
Table~\ref{tab:gmr_metrics} summarizes variable selection performance averaged across all simulation scenarios. The $\pi$-SGL penalty demonstrates consistently high recall at both the group and within-group levels, with group-level recall ranging from 0.77 to 1.00 and within-group recall from 0.69 to 0.99 across all configurations, indicating that truly relevant covariates are rarely missed. Precision is moderate, ranging from 0.51 to 0.74 at the group level and 0.44 to 0.76 at the within-group level, reflecting a tendency to retain some irrelevant covariates. 

\begin{table}[H]
    \centering
    \caption{Mean variable selection performance across simulation scenarios for Gaussian mixture regression with $\pi$-SGL penalty, averaged over 100 replications. Standard deviations are shown in parentheses.}
    \resizebox{\ifdim\width>\linewidth\linewidth\else\width\fi}{!}{
    \begin{tabular}{rllrlllllll}
    \toprule
    \multicolumn{4}{c}{ } & \multicolumn{3}{c}{Group selection} & \multicolumn{3}{c}{Within-group selection} & \multicolumn{1}{c}{ } \\
    \cmidrule(l{3pt}r{3pt}){5-7} \cmidrule(l{3pt}r{3pt}){8-10}
    p & $\sigma^2$ & $\pi$ & n & Precision & Recall & F1 & Precision & Recall & F1 & Direction acc.\\
    \midrule
    10 & equal & equal & 300 & 0.65 (0.07) & 0.97 (0.04) & 0.77 & 0.66 (0.08) & 0.96 (0.05) & 0.78 & 1.00\\
    10 & equal & unequal & 300 & 0.65 (0.08) & 0.98 (0.04) & 0.78 & 0.68 (0.08) & 0.95 (0.07) & 0.79 & 1.00\\
    10 & unequal & equal & 300 & 0.70 (0.10) & 0.93 (0.06) & 0.80 & 0.72 (0.08) & 0.87 (0.08) & 0.78 & 0.98\\
    10 & unequal & unequal & 300 & 0.73 (0.12) & 0.88 (0.14) & 0.78 & 0.73 (0.11) & 0.85 (0.15) & 0.77 & 0.99\\
    \addlinespace
    10 & equal & equal & 500 & 0.68 (0.06) & 1.00 (0.01) & 0.81 & 0.70 (0.07) & 0.99 (0.01) & 0.82 & 1.00\\
    10 & equal & unequal & 500 & 0.70 (0.08) & 1.00 (0.00) & 0.82 & 0.73 (0.09) & 0.99 (0.02) & 0.84 & 1.00\\
    10 & unequal & equal & 500 & 0.73 (0.08) & 0.97 (0.03) & 0.83 & 0.75 (0.08) & 0.92 (0.07) & 0.82 & 0.99\\
    10 & unequal & unequal & 500 & 0.74 (0.11) & 0.92 (0.15) & 0.80 & 0.76 (0.10) & 0.91 (0.14) & 0.82 & 1.00\\
    \addlinespace
    25 & equal & equal & 300 & 0.51 (0.07) & 0.94 (0.07) & 0.65 & 0.44 (0.05) & 0.87 (0.12) & 0.58 & 0.99\\
    25 & equal & unequal & 300 & 0.51 (0.08) & 0.89 (0.14) & 0.65 & 0.44 (0.05) & 0.81 (0.17) & 0.56 & 0.98\\
        25 & unequal & equal & 300 & 0.61 (0.16) & 0.81 (0.21) & 0.66 & 0.56 (0.17) & 0.70 (0.18) & 0.59 & 0.97\\
    25 & unequal & unequal & 300 & 0.63 (0.14) & 0.77 (0.28) & 0.64 & 0.57 (0.15) & 0.69 (0.26) & 0.58 & 0.96\\
    \addlinespace
    25 & equal & equal & 500 & 0.62 (0.14) & 0.98 (0.03) & 0.75 & 0.58 (0.15) & 0.96 (0.06) & 0.72 & 0.99\\
    25 & equal & unequal & 500 & 0.60 (0.11) & 0.98 (0.03) & 0.74 & 0.54 (0.11) & 0.95 (0.07) & 0.69 & 1.00\\
    25 & unequal & equal & 500 & 0.64 (0.18) & 0.92 (0.09) & 0.74 & 0.63 (0.18) & 0.87 (0.10) & 0.71 & 0.98\\
    25 & unequal & unequal & 500 & 0.68 (0.17) & 0.93 (0.10) & 0.77 & 0.63 (0.18) & 0.87 (0.13) & 0.71 & 0.98\\
    \bottomrule
    \end{tabular}}
    \label{tab:gmr_metrics}
\end{table}

\begin{figure}[H]
    \centering
    \caption{GMR simulation study: mean variable selection performance by number of covariates and sample size for Gaussian mixture regression, averaged across all simulation scenarios.}
    \includegraphics[width=1\linewidth]{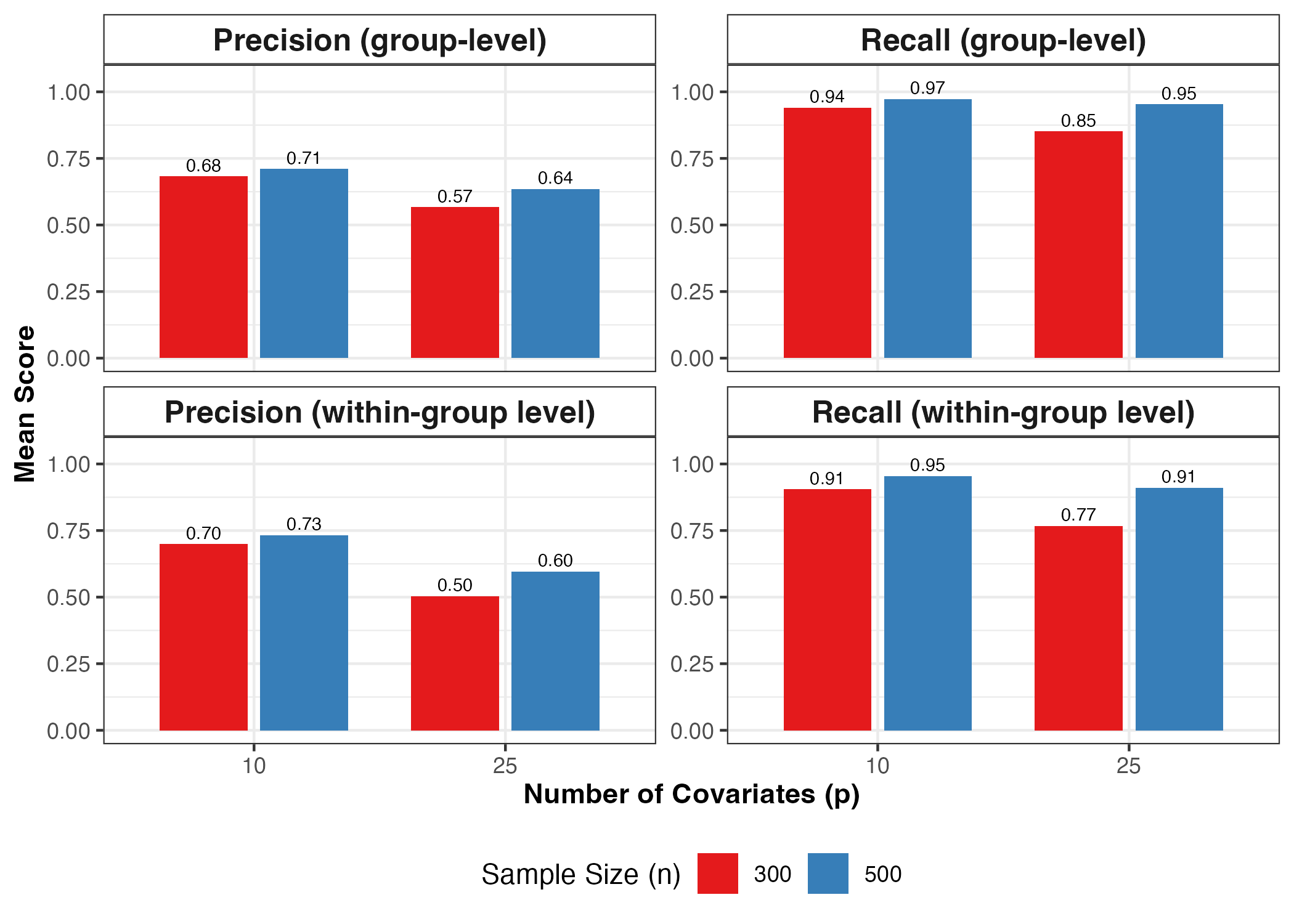}
    \label{fig:metrics_gmr}
\end{figure}

Performance improves with sample size across all metrics. As shown in Figure~\ref{fig:metrics_gmr}, increasing $n$ from 300 to 500 yields consistent gains in both precision and recall at the group and within-group levels, with the improvement most pronounced for $p=25$. Increasing the number of covariates from $p=10$ to $p=25$ reduces precision noticeably, which is expected as the higher-dimensional setting introduces more opportunities for false positives. Recall remains relatively robust to increasing $p$, particularly at $n=500$. Direction accuracy is near-perfect across all scenarios, indicating that when covariates are retained, their effect directions are correctly identified.

\begin{figure}[H]
    \centering
    \caption{GMR simulation study: Mean proportion of datasets selecting each mixture order $\hat{G}$ under BIC, averaged across all simulation scenarios. Results are stratified by true mixture order ($G = 3$, $G = 4$) and number of covariates ($p = 10$, $p = 25$). True order is highlighted in green.}
    \includegraphics[width=1\linewidth]{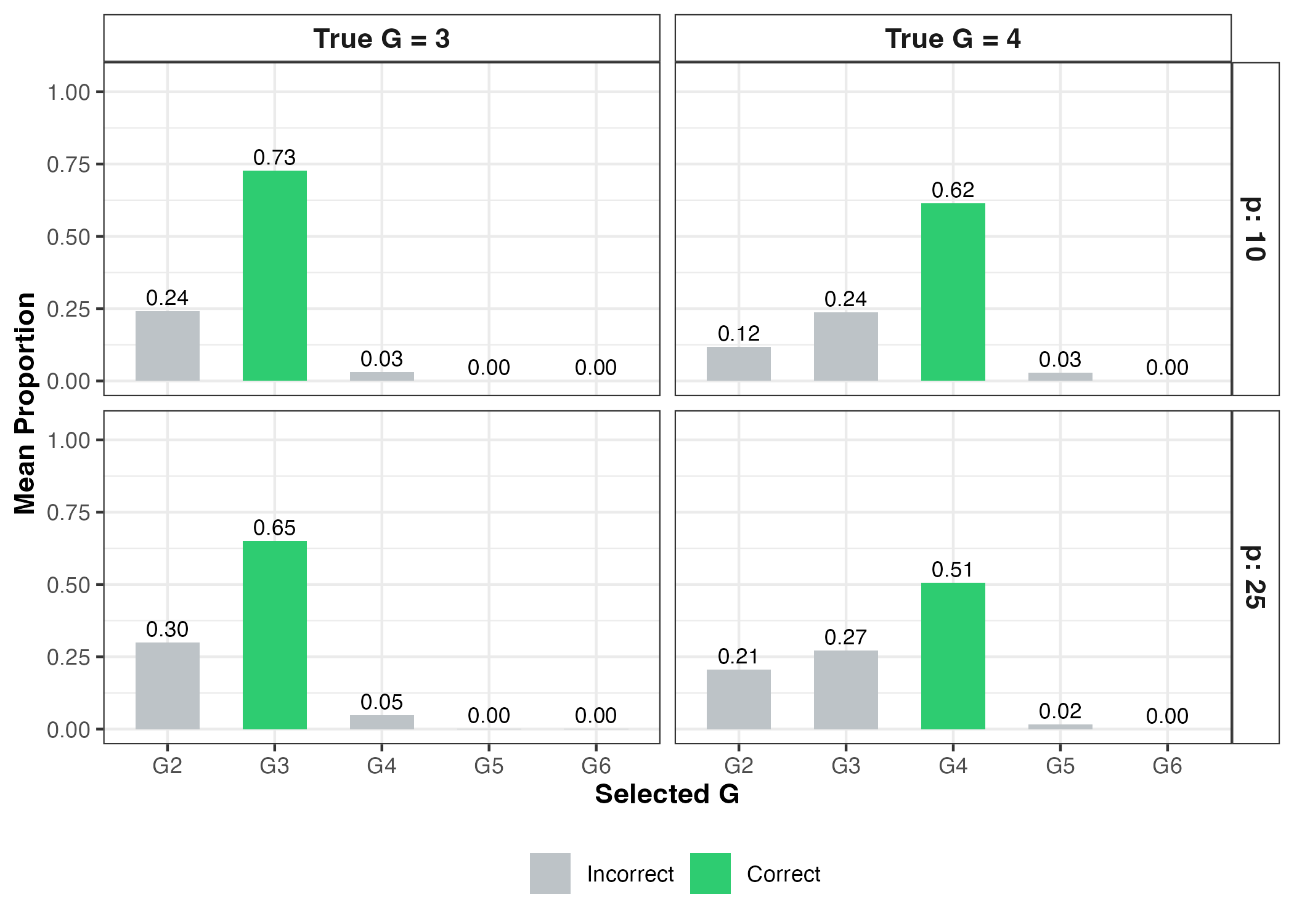}
    
    \label{fig:order_gmr}
\end{figure}

Model order selection via BIC is summarized in Figure~\ref{fig:order_gmr}. For $G=3$, BIC selects the correct order in 73\% of replications at $p=10$ and 65\% at $p=25$, with most errors attributable to under-selection of $\hat{G}=2$. Performance is weaker for $G=4$, with correct selection rates of 62\% at $p=10$ and 51\% at $p=25$, and errors are mostly attributed to under-selection of $\hat{G}= \{2, 3\}$. The difficulty in identifying $G=4$ components is consistent with the greater overlap between components in higher-order mixtures and represents a known limitation of BIC-based order selection in finite mixture models.

\subsubsection{Penalty comparison}

When mixing proportions are equal, the $\pi$-weighting in the SGL penalty has no effect, since all components receive the same weight. The comparison between weighted and unweighted SGL is therefore only informative under unequal mixing proportions, which we adopt as the setting for this comparison. In addition, as \citet{khalili2007} did not include a penalty on $\sigma^2$, we are interested in how the inclusion of such a penalty influences model fit. We consider four combinations of penalty functions: $\pi$-SGL with the $\sigma^2$ penalty, $\pi$-SGL without the $\sigma^2$ penalty, unweighted SGL with the $\sigma^2$ penalty, and unweighted SGL without the $\sigma^2$ penalty. These configurations were evaluated under the scenario: $n=500$, $G=3$, $p=10$, with unequal mixing proportions and unequal component variances, across 50 replications.
\begin{table}[H]
    \centering
    \caption{Mean variable selection performance across four penalty configurations for $G=3$, $n=500$, $p=10$ with unequal mixing proportions and unequal component variances, averaged over 50 replications of Gaussian mixture regression.}
    \begin{tabular}{llrrrrrrr}
    \toprule
    \multicolumn{2}{c}{ } & \multicolumn{3}{c}{Group selection} & \multicolumn{3}{c}{Within-group selection} & \multicolumn{1}{c}{ } \\
    \cmidrule(l{3pt}r{3pt}){3-5} \cmidrule(l{3pt}r{3pt}){6-8}
    $\sigma^2$ pen. & $\pi$-weighted & Prec. & Rec. & F1 & Prec. & Rec. & F1 & Direction acc.\\
    \midrule
    No & No & 0.64 & 0.97 & 0.77 & 0.62 & 0.88 & 0.70 & 0.92\\
    No & Yes & 0.63 & 0.99 & 0.77 & 0.65 & 0.91 & 0.75 & 0.96\\
    Yes & No & 0.74 & 0.97 & 0.84 & 0.75 & 0.90 & 0.81 & 0.99\\
    Yes & Yes & 0.67 & 0.98 & 0.80 & 0.71 & 0.90 & 0.79 & 0.99\\
    \bottomrule
    \end{tabular}
    \label{tab:penalty_metrics}
\end{table}

Table~\ref{tab:penalty_metrics} shows variable selection performance across four combinations of penalty functions. Group-level F1 scores range from 0.77 to 0.84 and within-group F1 scores from 0.70 to 0.81, with the inclusion of $\sigma_g^2$-penalty improving precision while leaving recall largely unchanged. Direction accuracy is consistently high across all configurations. These results suggest that variable selection is relatively robust to the choice of auxiliary penalties.

\begin{table}[H]
    \centering
    \caption{Mean proportion of replications selecting each mixture order $\hat{G}$ using BIC, across four penalty configurations with true order $G=3$.}
    \begin{tabular}{rllrrrrr}
    \toprule
    $\sigma^2$ pen. & $\pi$-weighted & $G_2$ & $G_3$ & $G_4$ & $G_5$ & $G_6$\\
    \midrule
    No & No & 0.10 & 0.05 & 0.19 & 0.23 & 0.16\\
    No & Yes & 0.07 & 0.21 & 0.28 & 0.22 & 0.14\\
    Yes & No & 0.42 & \textbf{0.52} & 0.06 & 0.00 & 0.00\\
    Yes & Yes & 0.24 & \textbf{0.72} & 0.04 & 0.00 & 0.00\\
    \bottomrule
    \end{tabular}

    \label{tab:penalty_order}
\end{table}

However, Table~\ref{tab:penalty_order} reveals a starkly different picture for model order selection. When the $\sigma^2$ penalty is included, the correct order ($\hat G = 3$) is selected in 52\% of replications without $\pi$-weighting and 72\% of replications with $\pi$-weighting, indicating that the $\pi$-weighted regularization penalty, in combination with the $\sigma^2$ penalty, improves order selection. Without the $\sigma^2$ penalty, the selected order spreads across $\hat G \in \{2,3,4,5,6\}$ with no dominant mode, selecting the correct order in only 21\% and 5\% of replications for the $\pi$-weighted and unweighted SGL configurations respectively. This demonstrates that the $\sigma^2$ penalty plays a critical role with respect to order selection, by preventing variance parameters from collapsing toward zero, consistent with the theoretical motivation of \citet{chenINFERENCENORMALMIXTURES2008}.

\subsection{Poisson mixture regression}
Data were generated from a $G$-component Poisson mixture regression (PMR) model with $p$ covariates. The design matrix was generated identically to the Gaussian case. Responses were generated as $y_i \sim \text{Poisson}(\exp(\mathbf{x}_i^\top \boldsymbol{\beta}_g))$ for the assigned component $g$, using the log link. The true coefficient matrix was constructed using the same sparsity structure as the Gaussian case, with nonzero coefficients drawn uniformly from $[0.3, 1.0]$ with random sign. Component intercepts were evenly spaced to ensure that expected counts $\exp(\beta_{0g})$ remained in a moderate range of $(0, 3]$. Mixing proportions were generated as described in Section~\ref{sec:gmr_sim}, with equal and unequal cases. Each scenario was replicated 100 times with fixed true parameters across replications. The simulation design is summarized in Table~\ref{tab:poisson_design}.

\begin{table}[H]
    \centering
    \caption{Poisson mixture regression simulation design. Each scenario was replicated 100 times with fixed true parameters across replications}
    \begin{tabular}{ll}
        \hline
         Simulation Parameter & Values\\
         \hline
         Number of components ($G$) & 2, 4 \\
         Number of covariates ($p$) & 10, 25\\
         Sample size ($n$) & 300, 500 \\
         Mixing proportions ($\boldsymbol{\pi}$) & Equal, Unequal\\
         Relevant covariate proportion ($\delta_p$) & 0.3, 0.5\\
         Relevant component proportion ($\delta_w$) & 0.3, 0.5\\
         \hline
    \end{tabular}
    \label{tab:poisson_design}
\end{table}

\begin{figure}[H]
    \centering
    \caption{PMR simulation study: mean variable selection performance by number of covariates and sample size, averaged across all simulation scenarios.}
    \includegraphics[width=1\linewidth]{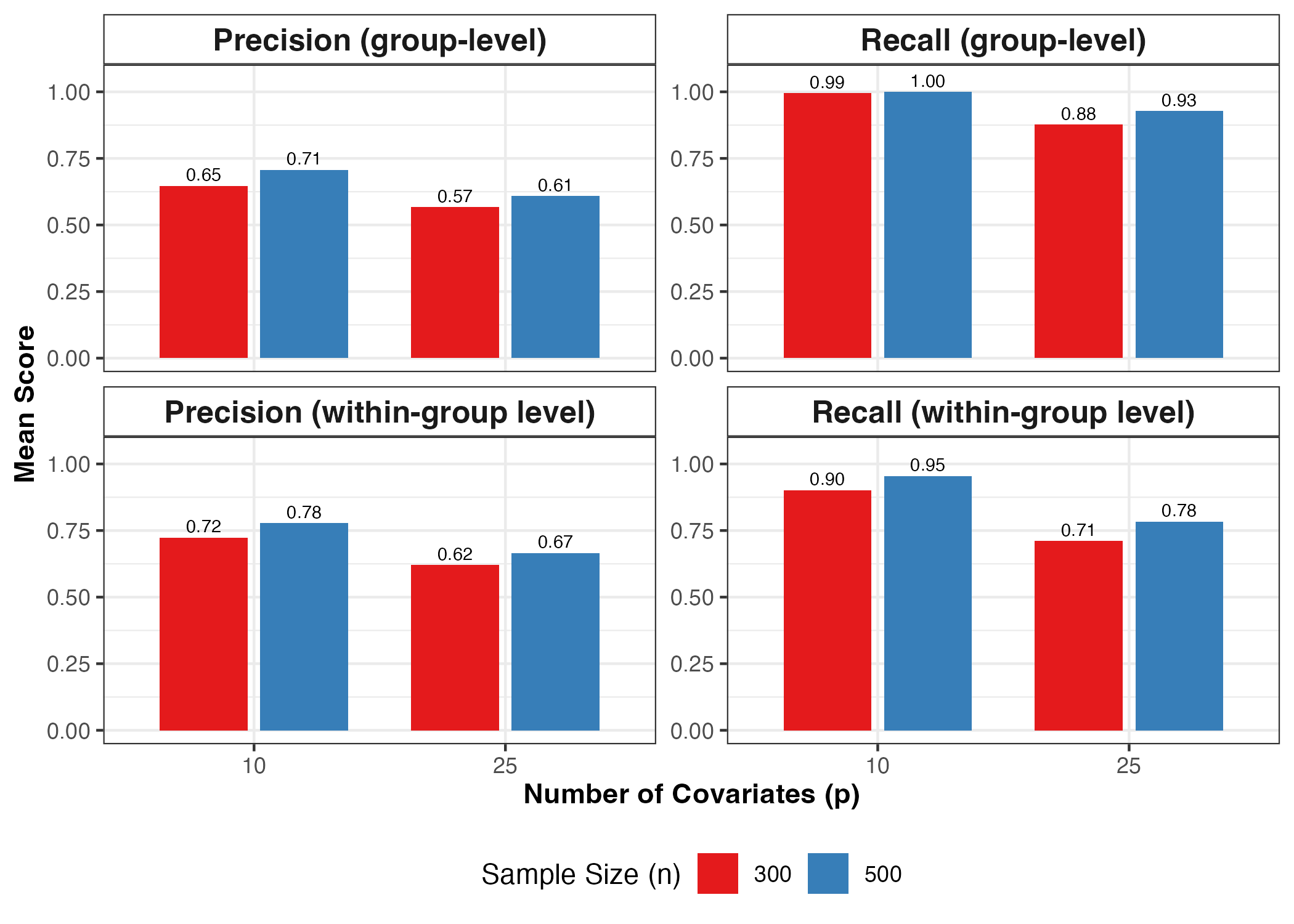}
    \label{fig:poisson_summary}
\end{figure}

Table~\ref{tab:poisson_results} summarizes variable selection performance for the Poisson mixture regression with $\pi$-SGL penalty. The fitted models maintain high recall at both group and within-group levels, group-level recall ranges from 0.88 to 1.00 and within-group recall from 0.71 to 0.95 across all combinations of simulation settings. Precision is moderate, ranging from 0.50 to 0.77 at the group level and slightly higher values of 0.56 to 0.84 at the within-group level. Direction accuracy is consistently high across all scenarios. As illustrated in Figure~\ref{fig:poisson_summary}, performance improves with sample size at both levels, and increasing $p$ from 10 to 25 reduces precision while recall remains relatively stable, consistent with the Gaussian findings.

\begin{table}[H]
    \centering
    \caption{Variable selection performance across simulation scenarios for Poisson mixture regression with $\pi$-SGL penalty, averaged over 100 replications. Standard deviations are shown in parentheses.}
    \resizebox{\ifdim\width>\linewidth\linewidth\else\width\fi}{!}{
    \begin{tabular}{rrllllllll}
    \toprule
    \multicolumn{3}{c}{ } & \multicolumn{3}{c}{Group selection} & \multicolumn{3}{c}{Within-group selection} & \multicolumn{1}{c}{ } \\
    \cmidrule(l{3pt}r{3pt}){4-6} \cmidrule(l{3pt}r{3pt}){7-9}
    p & n & $\pi$ & Precision & Recall & F1 & Precision & Recall & F1 & Direction acc.\\
    \midrule
    10 & 300 & equal & 0.71 (0.10) & 0.99 (0.01) & 0.83 & 0.79 (0.08) & 0.90 (0.12) & 0.84 & 0.93\\
    10 & 300 & unequal & 0.58 (0.08) & 0.99 (0.01) & 0.73 & 0.66 (0.07) & 0.90 (0.12) & 0.76 & 0.93\\
    10 & 500 & equal & 0.77 (0.10) & 1.00 (0.00) & 0.87 & 0.84 (0.07) & 0.95 (0.07) & 0.89 & 0.95\\
    10 & 500 & unequal & 0.64 (0.09) & 1.00 (0.00) & 0.78 & 0.72 (0.05) & 0.95 (0.07) & 0.82 & 0.95\\
    \addlinespace
    25 & 300 & equal & 0.63 (0.09) & 0.88 (0.15) & 0.72 & 0.69 (0.09) & 0.71 (0.30) & 0.67 & 0.91\\
    25 & 300 & unequal & 0.50 (0.07) & 0.88 (0.15) & 0.63 & 0.56 (0.07) & 0.71 (0.30) & 0.60 & 0.91\\
    25 & 500 & equal & 0.67 (0.09) & 0.93 (0.13) & 0.77 & 0.73 (0.07) & 0.78 (0.26) & 0.74 & 0.92\\
    25 & 500 & unequal & 0.55 (0.07) & 0.93 (0.13) & 0.68 & 0.60 (0.05) & 0.78 (0.26) & 0.67 & 0.92\\
    \bottomrule
    \end{tabular}}
    \label{tab:poisson_results}
\end{table}

\begin{figure}[H]
    \centering
    \caption{PMR simulation study: mean proportion of replications selecting each mixture order $\hat{G}$ under BIC, stratified by true mixture order ($G=2$, $G=4$) and number of covariates ($p=10$, $p=25$). True order is highlighted in green.}
    \includegraphics[width=1\linewidth]{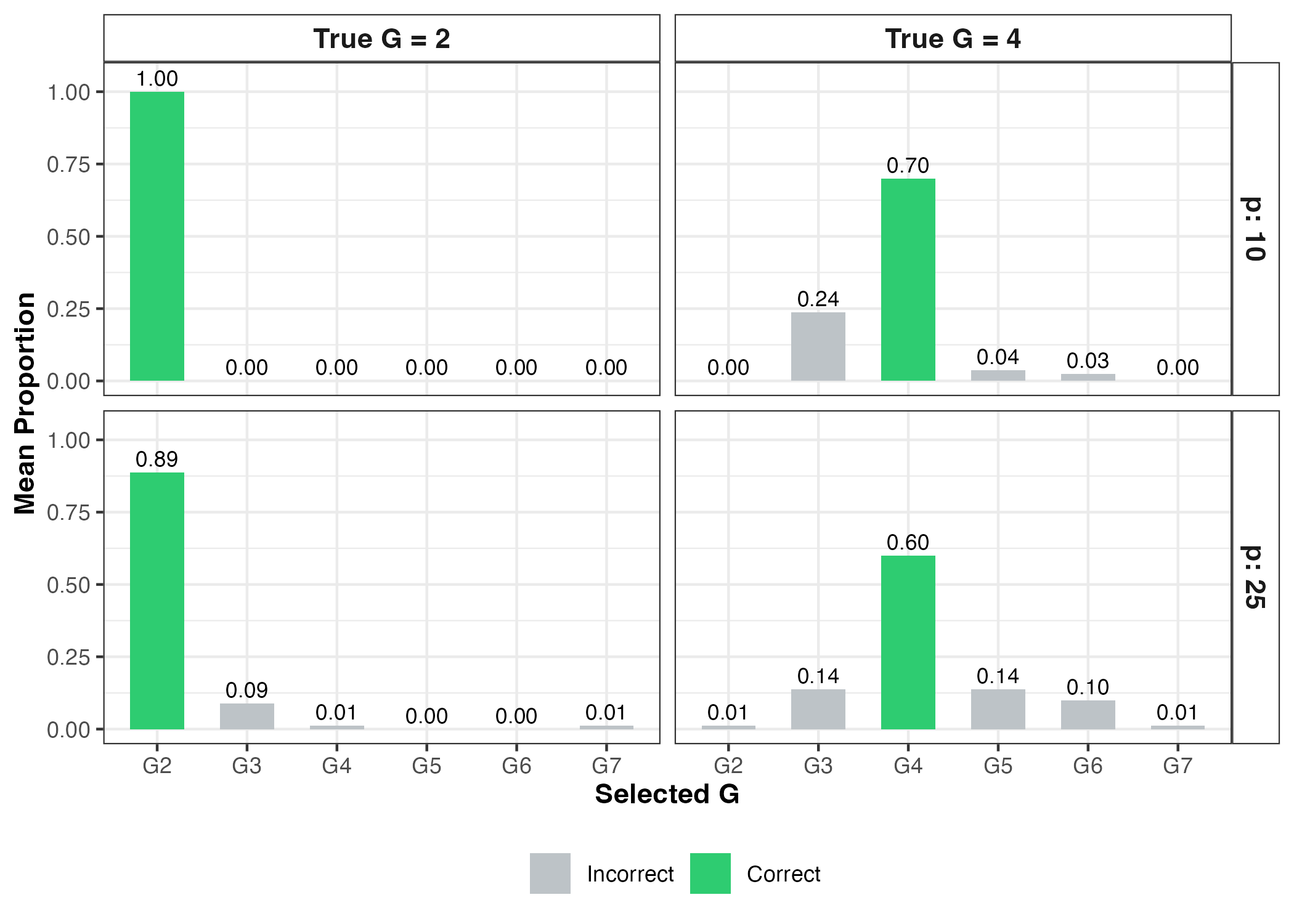}
    \label{fig:poisson_order}
\end{figure}

Order selection results are shown in Figure~\ref{fig:poisson_order}. For $G=2$, BIC selects the correct order in all replications at $p=10$ and 89\% at $p=25$. Performance is weaker for $G=4$, with correct selection rates of 70\% at $p=10$ and 60\% at $p=25$. The greater difficulty in identifying $G=4$ components is again consistent with the Gaussian findings and reflects the general challenge of distinguishing higher-order mixture components under moderate sample sizes. When $G=2,$ the order is set to the boundary of the smallest possible order in finite mixture models. So, the better performance for $G=2$ might be due to such boundary effect that the selected order can not be go lower than 2.

\section{Real data analysis}\label{sec:data}

To evaluate the practical utility of the proposed $\pi$-SGL penalized finite mixture regression framework, we apply it to two real datasets that reflect distinct data-generating contexts. The first is an ecological dataset on bat species morphology, where bat forearm length is modeled as a continuous response using a finite Gaussian mixture regression model. The second is a health economics dataset, where the response is the number of visits to a physician's office, motivating the use of a finite Poisson mixture regression model. Together, these applications demonstrate the flexibility of the $\pi$-SGL penalty across different FMR with distributions from the exponential family. In each case, the $\pi$-SGL penalty simultaneously performs variable selection at both the group and within-group levels, and BIC is used to select the tuning parameters and the number of mixture components $G$ from a candidate set of $\{2, 3, 4, 5, 6, 7\}$.

\subsection{Chiroptera data}
The dataset used in this analysis is sourced from PanTHERIA, a global species-level database compiling life-history, geographical, and ecological traits of extant and recently extinct mammalian species~\citep{jonesPanTHERIASpecieslevelDatabase2009}. From this database, observations on 589 Chiroptera species were extracted, with bat forearm length (mm) serving as the response variable. Seven explanatory variables were considered: absolute latitude, human population density, population area, average monthly precipitation, mean temperature, average evapotranspiration rate (AET), and adult body mass. Bat forearm length is known to vary substantially across species in response to differences in ecology, geographic range, and thermoregulatory demands, suggesting that a single regression model may be insufficient to capture the heterogeneous relationships between forearm length and environmental covariates. A finite mixture regression model is therefore appropriate, as it allows the population of bat species to be clustered into latent subpopulations with distinct covariate effects. Prior to model fitting, body mass, population density, and area were log-transformed to address right skewness and the presence of outliers; a $\log(\text{population density} + 0.01)$ transformation was applied to accommodate zero values. Absolute latitude was used to account for symmetry about the equator. All covariates were subsequently standardized to have mean zero and unit variance. 

To assess the sensitivity of the results to the homogeneity assumption on the residual variances, two versions of the GMR model with $\pi$-SGL were fitted to the Chiroptera dataset: one allowing for component-specific variances and one imposing a common variance across all components, with the results of both models summarized in Tables~\ref{tab:bat_unequal} and \ref{tab:bat_equal} respectively. Under a common variance model, a penalty on $\sigma_g^2$ is not required during model fitting.

The unequal variance model identified three latent subpopulations among the 589 bat species, with component 2 comprising the largest sample size ($n = 300$), followed by component 1 ($n = 205$) and component 3 ($n = 84$). The estimated residual standard deviations were similar for components 1 and 2 ($\hat{\sigma} \approx 4.8$) but considerably larger for component 3 ($\hat{\sigma} = 11.46$), indicating greater variability in forearm length among species assigned to that subpopulation. $\log(\text{Body Mass})$ was retained across all three components with estimated coefficients of 7.14, 16.06 and 29.21, reflecting an increasingly strong association between body mass and forearm length across subpopulations. $\log(\text{Pop.\ Density})$, Temperature, and $\log(\text{Area})$ were retained only in components 1 and/or 2, with Temperature exhibiting opposing directions across those two components ($-0.15$ and $0.56$, respectively). Precipitation, absolute latitude, and AET were excluded entirely across all three components.

\begin{table}[H]
\centering
\caption{Estimated parameters from $\pi$-SGL-GMR fitted to the Chiroptera dataset ($\lambda = 6.91$, $\alpha = 0.9$, $\text{BIC} = 4166.17$). Zero coefficients (red) indicate variables excluded by the penalty.}
\begin{tabular}{l r r r}
\hline
Parameter & Comp. 1 & Comp. 2 & Comp. 3  \\
&  ($n=205$) &  ($n=300$) &  ($n=84$) \\
\hline
Intercept           & 42.91 & 51.59 & 54.23 \\
$\log(\text{Body Mass})$     & 7.14  & 16.06 & 29.21 \\
$\log(\text{Pop.\ Density})$ & 1.13  & \textcolor{red}{0.00} & \textcolor{red}{0.00} \\
Temperature         & $-$0.15 & 0.56  & \textcolor{red}{0.00} \\
$\log(\text{Area})$ & $-$0.24 & \textcolor{red}{0.00} & \textcolor{red}{0.00} \\
Precipitation       & \textcolor{red}{0.00} & \textcolor{red}{0.00} & \textcolor{red}{0.00} \\
$\text{abs}(\text{Latitude})$ & \textcolor{red}{0.00} & \textcolor{red}{0.00} & \textcolor{red}{0.00} \\
AET                 & \textcolor{red}{0.00} & \textcolor{red}{0.00} & \textcolor{red}{0.00} \\
\hline
$\hat{\sigma}$      & 4.77  & 4.78  & 11.46 \\
\hline
\end{tabular}
\label{tab:bat_unequal}
\end{table}

Under the equal variance constraint, three subpopulations were again recovered: a dominant central component ($n = 428$) flanked by two smaller components ($n = 81$ and $n = 80$). The shared residual standard deviation was estimated at $\hat{\sigma} = 6.29$. $\log(\text{Body Mass})$ again emerged as the most influential covariate across all components, with coefficients of comparable magnitude to the unequal variance model. Notably, the equal variance model retained additional covariates within components: Precipitation appeared in components 2 and 3, Temperature and $\log(\text{Area})$ were retained in component 2, and AET in component 3.

\begin{table}[H]
\centering
\caption{Estimated parameters from $\pi$-SGL-GMR with equal variances fitted to 
the Chiroptera dataset ($\lambda = 5.63$, $\alpha = 0.8$, $\text{BIC} = 4290.87$). Zero coefficients (red) indicate variables excluded by the penalty.}
\begin{tabular}{l r r r}
\hline
Parameter & Comp. 1 & Comp. 2 & Comp. 3 \\
& ($n=81$) & ($n=428$) & ($n=80$) \\
\hline
Intercept           & 41.82 & 50.04 & 49.85 \\
$\log(\text{Body Mass})$     & 4.31  & 15.09 & 31.77 \\
$\log(\text{Pop.\ Density})$ & 0.96  & 0.27  & 0.51  \\
Temperature         & \textcolor{red}{0.00} & 0.10  & \textcolor{red}{0.00}  \\
$\log(\text{Area})$ & \textcolor{red}{0.00} & $-$0.23 & \textcolor{red}{0.00} \\
Precipitation       & \textcolor{red}{0.00} & 0.29 & 4.40  \\
$\text{abs}(\text{Latitude})$ & \textcolor{red}{0.00} & \textcolor{red}{0.00} & \textcolor{red}{0.00} \\
AET                 & \textcolor{red}{0.00} & \textcolor{red}{0.00} & $-$6.15 \\
\hline
$\hat{\sigma}$      & \multicolumn{3}{c}{6.29} \\
\hline
\end{tabular}
\label{tab:bat_equal}
\end{table}

Comparing the two models, both agree on the three-subpopulation structure and on the centrality of $\log(\text{Body Mass})$ as the primary predictor of forearm length. However, the variance assumption meaningfully influences which covariates are selected and where. A clear example is $\log(\text{Pop.\ Density})$: under unequal variances it is retained only in component 1, whereas the equal variance model retains it across all three components with similar small coefficients. More broadly, the equal variance constraint appears to redistribute unexplained heterogeneity from the residual variance into the covariate structure, resulting in more covariates being retained within components. The unequal variance model, favored by BIC, achieves a more parsimonious fit by allowing component-specific variances to absorb some of this variability.

In both models the selected penalty placed most of its weight on the $\pi$-weighted LASSO component ($\alpha = 0.9$ for unequal variance model and $\alpha = 0.8$ for equal variance model), with a small contribution from the group LASSO term. In the unequal variance model, Precipitation, absolute latitude, and AET were excluded across all three components, while in the equal variance model only absolute latitude was excluded everywhere, with AET retained in component 3 and Precipitation retained in components 2 and 3. The active group LASSO term encourages covariates to be dropped simultaneously across components, while the dominant LASSO term governs component-specific exclusions; together they yield covariate-level sparsity where a variable is weak across all subpopulations and component-level sparsity elsewhere. 

\subsection{Elderly patient visits}
The data used in this analysis are drawn from the 1987 National Medical Expenditure Survey (NMES), a nationally representative sample of the civilian non-institutionalized population of the United States \citep{deb1997}. The subsample considered here consists of 4406 individuals aged 66 and over, all of whom are covered by Medicare. The response variable is the number of visits to a physician in an office setting. The explanatory variables include measures of health such as self-perceived health status, number of chronic conditions, and activities of daily living (ADL) difficulty. Sociodemographic variables include age, years of schooling, family income, gender, race (whether they are African American or not), marital status, and employment status. Regional indicators (midwest, northeast, west), private insurance coverage and Medicaid enrollment are also included. Prior to model fitting, the health status variable was dummy-encoded with average health as the reference category, yielding indicators for poor and excellent health. The ADL variable was dummy-encoded with normal ADL status as the reference, and region was dummy-encoded with other as the reference category, yielding indicators for northeast, midwest, and west. All covariates were subsequently standardized to have mean zero and unit variance.

\begin{table}[H]
    \centering
    \caption{Estimated parameters from $\pi$-SGL-PMR fitted to the NMES dataset ($\lambda = 42.59$, $\alpha = 1$, $\text{BIC} = 24832.19$). Zero coefficients (red) indicate variables excluded by 
the penalty.}
    \begin{tabular}{l c c c c}
        \hline
        Parameter & Comp 1  & Comp 2  & Comp 3  & Comp 4  \\
         & ($n=64$) &  ($n=2174$) &  ($n=705$) & ($n=1463$) \\
        \hline
        Intercept   & 3.976 & 1.416 & 2.410 & -0.500 \\
        Age         & 0.022 & \textcolor{red}{0.000} & 0.021 & \textcolor{red}{0.000} \\
        School      & 0.168 & 0.109 & 0.082 & 0.167 \\
        Income      & -0.085 & -0.013 & -0.014 & 0.008 \\
        Chronic     & 0.001 & 0.238 & 0.101 & 0.462 \\
        Health (poor) & 0.431 & 0.224 & 0.291 & \textcolor{red}{0.000} \\
        Health (excel.) & 0.475 & -0.237 & -0.237 & \textcolor{red}{0.000} \\
        ADL (limited)  & 0.079 & 0.044 & 0.032 & \textcolor{red}{0.000} \\
        Region (midwest) & 0.014 & 0.039 & 0.042 & \textcolor{red}{0.000} \\
        Region (northeast) & 0.290 & \textcolor{red}{0.000} & 0.073 & \textcolor{red}{0.000} \\
        Region (west) & 0.061 & 0.065 & 0.121 & \textcolor{red}{0.000} \\
        African American & -0.029 & -0.033 & \textcolor{red}{0.000} & -0.241 \\
        Gender (male) & 0.142 & -0.098 & 0.020 & -0.147 \\
        Married     & -0.298 & -0.001 & -0.060 & 0.004 \\
        Employed    & 0.150 & -0.018 & \textcolor{red}{0.000} & \textcolor{red}{0.000} \\
        Insurance   & -0.422 & 0.239 & 0.139 & 0.727 \\
        Medicaid    & -0.386 & 0.189 & 0.092 & \textcolor{red}{0.000} \\
        \hline
    \end{tabular}
    \label{tab:nmes_results}
\end{table}

 As described in Table~\ref{tab:nmes_results}, the $\pi$-SGL penalized finite Poisson mixture model identified four latent subpopulations among the 4406 elderly Medicare recipients, selected via BIC ($\lambda = 42.59$, $\alpha = 1$, $\text{BIC} = 24832.19$). Component 2 comprises the largest subpopulation ($n = 2174$), followed by component 4 ($n = 1463$), component 3 ($n = 705$), and component 1 ($n = 64$). The intercepts vary across components (3.976, 1.416, 2.410, and -0.500 respectively), suggesting differences in baseline physician visit rates across latent subpopulations.

 Tuning parameter $\alpha = 1$ was selected, indicating that variable selection was driven entirely by the $\pi$-weighted LASSO component of the penalty. No covariate was completely excluded across all components; every variable was retained in at least one component, suggesting that each covariate carries some association with physician visit counts in at least one latent subpopulation. Component 4, despite being the second largest component, exhibits the sparsest covariate structure, with only School, Chronic, Insurance, African American, Income, Married and Gender retained. The dominant covariate in this component is Insurance (0.727), followed by Chronic (0.462), suggesting that within this subpopulation physician utilization is primarily associated with insurance coverage and chronic disease burden. In contrast, components 1, 2, and 3 retain a broader set of covariates. Chronic conditions are retained across all four components, with coefficients of varying magnitudes (0.001, 0.238, 0.101, 0.462) pointing to heterogeneous associations between chronic disease burden and utilization across subpopulations. The positive coefficients indicate that a higher number of chronic conditions is associated with more physician visits.

The health status indicators exhibit a notable pattern. Health (poor) carries positive coefficients in components 1, 2, and 3 (0.431, 0.224, and 0.291), while Health (excellent) is positive in component 1 (0.475), it is negative in components 2 and 3 ($-0.237$ in both). Both poor and excellent health having zero coefficients in component 4 suggest no association between health status and number of visits to the physician's office within this subpopulation. The positive coefficient on Health (excellent) in component 1 is somewhat counterintuitive; together with the negative coefficient on Insurance in that component ($-0.422$), it may reflect the small size of that component ($n = 64$) or residual confounding with other covariates. Insurance coverage is otherwise positively associated with physician visits in components 2, 3, and 4.

\section{Discussion}\label{sec:discussion}
This paper proposed a $\pi$-weighted Sparse Group LASSO ($\pi$-SGL) penalty for simultaneous group-level and within-group level variable selection in finite mixture regression models. The penalty is embedded within a Majorization-Minimization framework for the class of generalized linear models via an IRLS-based surrogate construction. 

Simulation studies across a range of scenarios demonstrated that the $\pi$-SGL penalized estimator performs well in recovering the true sparsity structure of both Gaussian and Poisson finite mixture regression models. At the group level, the method reliably identifies covariates that are irrelevant across all components, while at the within-group level it captures component-specific sparsity. Variable selection performance improves with increasing sample size, and performance is more reliable when the true number of components $G$ is correctly specified. The use of BIC for joint selection of $G$, $\lambda$, and $\alpha$ is motivated by the consistency results of \citet{keribin2000consistent}, though formal verification of these conditions in the $\pi$-SGL setting remains an open problem. Model order selection was observed to be more challenging for larger $G$ and unequal mixing proportions, consistent with the known difficulty of distinguishing minority components when observations are scarce.

Applied to the Chiroptera dataset, the method identified a multi-component Gaussian mixture regression model in which body mass emerged as the dominant covariate of bat forearm length, with covariate effects varying substantially across subpopulations. The two largest components captured the main variation in forearm length driven by morphological and environmental covariates, while smaller components clustered species primarily on the basis of their forearm length distribution rather than covariate effects. Applied to the elderly patient visits data under a Poisson mixture regression model, the method identified latent subpopulations with heterogeneous healthcare utilization patterns, with several covariates zeroed out within specific components and no covariate was completely removed across all components. One might suggest the use of sparse LASSO penalty is sufficient for variable selection problem in FMR. However, this might preclude the possibility of retaining a completely irrelevant covariate in the model as sparse LASSO penalty alone might not be sufficiently effective in this case. We note that \citet{deb1997} analyzed the same physician office visits response using a finite mixture of negative binomial regression models, motivated by the overdispersion present in the data; however, the variable selection problem was not considered in their work.

The principal theoretical limitation of the present work is the absence of formal asymptotic guarantees. We expect the penalized estimator to attain the usual $\sqrt{n}$ estimation rate under regularity conditions analogous to those of \citet{khalili2007}. Sign consistency at both the group and within-group levels may not be achievable for the plain $\pi$-SGL penalty without adaptive reweighting. This parallels two established settings: the non-adaptive sparse-group LASSO, for which the oracle property fails and is recovered only under adaptively weighted penalties \citep{poignardAsymptoticTheoryAdaptive2020}; and the $\ell_1$-penalized finite mixture model, in which simultaneous sparsity and $\sqrt{n}$-consistency cannot be attained without a two-stage adaptive procedure \citep{stadler1penalizationMixtureRegression}. The rate conditions required for $\sqrt{n}$-consistency and for sign consistency cannot be satisfied at once by a single non-adaptive penalty \citep{khalili2007, zouAdaptiveLassoIts2006}. The $\pi$-weights used in our penalty function are structural rather than adaptive in this sense: they rescale the group penalties by the mixing proportions, but are not random weights derived from a preliminary $\sqrt{n}$-consistent estimator that diverge on the truly zero coefficients, and so do not confer these properties. Establishing the asymptotic theory of $\pi$-SGL, and determining whether an adaptively weighted variant attains the oracle property, is left for future work. A second, practical limitation is computational: the per-iteration cost of the MM algorithm scales with $n$, $G$, and $p$, and can become burdensome as dimensionality grows. The method has been evaluated primarily in moderate-$p$ settings; its behavior in high-dimensional regimes where $p \gg n$ has not been characterized, though the non-asymptotic oracle inequalities developed for $\ell_1$-penalized mixtures \citep{stadler1penalizationMixtureRegression} offer a natural starting point.

A practical observation concerns the construction of $\lambda_{\max}$, the start of the regularization path. Computing $\lambda_{\max}$ in closed form for the $\pi$-SGL penalty is less straightforward than in the linear LASSO or sparse group LASSO, where the absence of a mixture structure yields a clean expression. We instead derived $\lambda_{\max}$ from the MM surrogate evaluated at initialization (see Appendix~\ref{app:lambda}), taking the minimizing endpoint of the $\alpha$-dependent penalty denominator to obtain a single value valid across the full $\alpha$ grid. This yields a conservative upper bound on the true $\lambda_{\max}$, defined as the smallest $\lambda$ that zeros all penalized coefficients. In our simulations the derived value consistently overshot this threshold; as little as one-half of the computed $\lambda_{\max}$ was sufficient to produce a fully sparse solution, so the upper portion of the path contributed only all-zero models. As a practical heuristic, the path can be started at a fraction of the computed $\lambda_{\max}$ to determine if the all-zero end of the solution path is preserved. If it is, this allows the grid to be concentrated on the range where covariates actually enter the model; otherwise, the starting fraction should be increased, or the computed $\lambda_{\max}$ used instead. A tighter, $\alpha$-specific bound that avoids the $\pi_{\min}$ relaxation is a natural refinement.

Several directions warrant further investigation. Extending the simulation study and methodology to high-dimensional settings with $p \gg n$ represents a natural and practically important generalization. As our MM algorithm separates the optimization across components, a coordinate descent approach to finding solutions might be possible. To facilitate adoption, reproducibility and its application in real data analyses, it is worth implementing a regularized finite mixture regression framework for distributions from the exponential family as a publicly available, user-friendly \texttt{R} software package.

\section{Disclosure statement}\label{disclosure-statement}

The authors have no conflicts of interest to declare.

\section{Data Availability Statement}\label{data-availability-statement}

Both datasets analyzed in this paper are publicly available. The Chiroptera morphology data were extracted from the PanTHERIA database
\citep{jonesPanTHERIASpecieslevelDatabase2009}, available at \url{https://doi.org/10.1890/08-1494.1}. The physician
office visits data are the \texttt{NMES1988} dataset distributed with the \texttt{R} package \texttt{AER} \citep{kleiberAER2008}, derived from the 1987--88 U.S.\ National Medical Expenditure Survey and previously analyzed by \citet{deb1997}. The $\pi$-SGL estimator is implemented in an \texttt{R} package available at \url{https://github.com/vjoshy/regMR}.

\section*{Author contributions}
\begin{enumerate}
    \item Vinay Joshy: Writing - original draft, methodology, formal analysis, software (\url{https://orcid.org/0009-0003-7083-1256})
    \item Dr. Zeny Feng: Writing – review \& editing, methodology, supervision (\url{https://orcid.org/0000-0002-8113-3619})
    \item Grace Stelter: Writing – review \& editing, methodology
    \item Dr. Lorna E. Deeth: Writing – review \& editing
    \item Dr. Alysha Cooper: Writing – review \& editing
\end{enumerate}

\bibliography{jcgs/ref.bib}

@book{mclachlanFiniteMixtureModels2000,
  title = {Finite Mixture Models},
  author = {McLachlan, Geoffrey and Peel, David},
  year = 2000,
  month = sep,
  series = {Wiley Series in Probability and Statistics},
  edition = {1},
  publisher = {Wiley},
  issn = {1940-6347},
  doi = {10.1002/0471721182},
  urldate = {2025-05-01},
  copyright = {http://doi.wiley.com/10.1002/tdm\_license\_1.1},
  isbn = {978-0-471-00626-8 978-0-471-72118-5}
}

@article{lerouxConsistentEstimationMixing1992,
  title = {Consistent Estimation of a Mixing Distribution},
  author = {Leroux, Brian G.},
  year = 1992,
  month = sep,
  journal = {The Annals of Statistics},
  volume = {20},
  number = {3},
  issn = {0090-5364},
  doi = {10.1214/aos/1176348772},
  urldate = {2026-05-27},
  langid = {english}
}

@article{rednerMixtureDensitiesMaximum1984,
  title = {Mixture Densities, Maximum Likelihood and the {{EM}} Algorithm},
  author = {Redner, Richard A. and Walker, Homer F.},
  year = 1984,
  month = apr,
  journal = {SIAM Review},
  volume = {26},
  number = {2},
  pages = {195--239},
  issn = {0036-1445, 1095-7200},
  doi = {10.1137/1026034},
  urldate = {2026-05-27},
  langid = {english}
}

@book{titteringtonStatisticalAnalysisFinite1985a,
  title = {Statistical Analysis of Finite Mixture Distributions},
  author = {Titterington, D. M. and Smith, Adrian F. M. and Makov, U. E.},
  year = 1985,
  series = {Wiley Series in Probability and Mathematical Statistics},
  publisher = {Wiley},
  address = {Chichester ; New York},
  isbn = {978-0-471-90763-3},
  lccn = {QA276.7 .T57 1985}
}

@article{dempsterMaximumLikelihoodIncomplete1977,
  title = {Maximum Likelihood from Incomplete Data Via the \emph{{EM}} Algorithm},
  author = {Dempster, A. P. and Laird, N. M. and Rubin, D. B.},
  year = 1977,
  month = sep,
  journal = {Journal of the Royal Statistical Society Series B: Statistical Methodology},
  volume = {39},
  number = {1},
  pages = {1--22},
  issn = {1369-7412, 1467-9868},
  doi = {10.1111/j.2517-6161.1977.tb01600.x},
  urldate = {2026-05-27},
  copyright = {https://academic.oup.com/journals/pages/open\_access/funder\_policies/chorus/standard\_publication\_model},
  langid = {english}
}

@article{wedelMixtureLikelihoodApproach1995,
  title = {A Mixture Likelihood Approach for Generalized Linear Models},
  author = {Wedel, Michel and DeSarbo, Wayne S.},
  year = 1995,
  month = mar,
  journal = {Journal of Classification},
  volume = {12},
  number = {1},
  pages = {21--55},
  issn = {0176-4268, 1432-1343},
  doi = {10.1007/BF01202266},
  urldate = {2026-05-27},
  copyright = {http://www.springer.com/tdm},
  langid = {english}
}

@book{wedelMarketSegmentation2000,
  title = {Market Segmentation},
  author = {Wedel, Michel and Kamakura, Wagner A.},
  year = 2000,
  series = {International Series in Quantitative Marketing},
  volume = {8},
  publisher = {Springer US},
  address = {Boston, MA},
  doi = {10.1007/978-1-4615-4651-1},
  urldate = {2026-05-27},
  copyright = {http://www.springer.com/tdm},
  isbn = {978-1-4613-7104-5 978-1-4615-4651-1}
}

@article{chenINFERENCENORMALMIXTURES2008,
  title = {Inference for Normal Mixtures in Mean and Variance},
  author = {Chen, Jiahua and Tan, Xianming and Zhang, Runchu},
  year = {2008},
  journal = {Statistica Sinica},
  volume = {18},
  number = {2},
  pages = {443--465},
  publisher = {Statistica Sinica},
  url = {http://www.jstor.org/stable/24308490}
}

@article{hunterTutorialMMAlgorithms2004a,
  title = {A Tutorial on {{MM}} Algorithms},
  author = {Hunter, David R and Lange, Kenneth},
  year = 2004,
  month = feb,
  journal = {The American Statistician},
  volume = {58},
  number = {1},
  pages = {30--37},
  publisher = {Informa UK Limited},
  issn = {0003-1305, 1537-2731},
  doi = {10.1198/0003130042836},
  urldate = {2025-07-12},
  langid = {english}
}

@article{schwarzEstimatingDimensionModel1978,
  title = {Estimating the Dimension of a Model},
  author = {Schwarz, Gideon},
  year = 1978,
  month = mar,
  journal = {The Annals of Statistics},
  volume = {6},
  number = {2},
  issn = {0090-5364},
  doi = {10.1214/aos/1176344136},
  urldate = {2026-05-22},
  langid = {english}
}

@article{keribin2000consistent,
  title = {Consistent Estimation of the Order of Mixture Models},
  journal = {Sankhy{\~{a}}: The Indian Journal of Statistics, Series A},
  volume = {62},
  year = 2000,
  month = feb,
  number = 1,
  pages = {49--66},
  author = {Keribin, C},
  langid = {english}
}

@article{friedmanRegularizationPathsGeneralized2010,
  title = {Regularization Paths for Generalized Linear Models via Coordinate Descent},
  author = {Friedman, Jerome and Hastie, Trevor and Tibshirani, Robert},
  year = 2010,
  journal = {Journal of Statistical Software},
  volume = {33},
  number = {1},
  issn = {1548-7660},
  doi = {10.18637/jss.v033.i01},
  urldate = {2025-04-29},
  langid = {english}
}

@article{khalili2007,
  title = {Variable Selection in Finite Mixture of Regression Models},
  author = {Khalili, Abbas and Chen, Jiahua},
  year = 2007,
  month = sep,
  journal = {Journal of the American Statistical Association},
  volume = {102},
  number = {479},
  pages = {1025--1038},
  issn = {0162-1459, 1537-274X},
  doi = {10.1198/016214507000000590},
  urldate = {2025-04-29},
  langid = {english}
}

@article{stadler1penalizationMixtureRegression,
  title = {{$\ell$}1-Penalization for Mixture Regression Models},
  author = {St{\"a}dler, Nicolas and B{\"u}hlmann, Peter and Van~De Geer, Sara},
  year = 2010,
  month = aug,
  journal = {TEST},
  volume = {19},
  number = {2},
  pages = {209--256},
  issn = {1133-0686, 1863-8260},
  doi = {10.1007/s11749-010-0197-z},
  urldate = {2026-07-20},
  copyright = {http://www.springer.com/tdm},
  langid = {english}
}

@article{deb1997,
  title = {Demand for Medical Care by the Elderly: A Finite Mixture Approach},
  shorttitle = {DEMAND FOR MEDICAL CARE BY THE ELDERLY},
  author = {Deb, Partha and Trivedi, Pravin K.},
  year = 1997,
  month = may,
  journal = {Journal of Applied Econometrics},
  volume = {12},
  number = {3},
  pages = {313--336},
  issn = {0883-7252, 1099-1255},
  doi = {10.1002/(SICI)1099-1255(199705)12:3<313::AID-JAE440>3.0.CO;2-G},
  urldate = {2026-05-19},
  copyright = {http://doi.wiley.com/10.1002/tdm\_license\_1.1},
  langid = {english}
}

@article{simonSparseGroupLasso2013,
  title = {A Sparse-Group Lasso},
  author = {Simon, Noah and Friedman, Jerome and Hastie, Trevor and Tibshirani, Robert},
  year = 2013,
  month = jan,
  journal = {Journal of Computational and Graphical Statistics},
  volume = {22},
  number = {2},
  pages = {231--245},
  issn = {1061-8600, 1537-2715},
  doi = {10.1080/10618600.2012.681250},
  urldate = {2025-07-04},
  langid = {english}
}

@article{cooperDominatingHyperplaneRegularization2025,
  title = {Dominating Hyperplane Regularization for Variable Selection in Multivariate Count Regression},
  author = {Cooper, Alysha and Feng, Zeny and Ali, Ayesha and Arciszewski, Tim and Deeth, Lorna},
  year = 2025,
  month = apr,
  eprint = {2504.05034},
  primaryclass = {stat},
  publisher = {arXiv},
  doi = {10.48550/arXiv.2504.05034},
  urldate = {2025-07-04},
  URL = {https://arxiv.org/abs/2504.05034},
  archiveprefix = {arXiv},
  langid = {english}
  }

@article{tibshirani1996,
 ISSN = {00359246},
 URL = {http://www.jstor.org/stable/2346178},
 author = {Robert Tibshirani},
 journal = {Journal of the Royal Statistical Society. Series B (Methodological)},
 number = {1},
 pages = {267--288},
 publisher = {[Royal Statistical Society, Oxford University Press]},
 title = {Regression Shrinkage and Selection via the Lasso},
 urldate = {2026-05-28},
 volume = {58},
 year = {1996}
}

@article{yuanModelSelectionEstimation2006a,
  title = {Model Selection and Estimation in Regression with Grouped Variables},
  author = {Yuan, Ming and Lin, Yi},
  year = 2006,
  month = feb,
  journal = {Journal of the Royal Statistical Society Series B: Statistical Methodology},
  volume = {68},
  number = {1},
  pages = {49--67},
  issn = {1369-7412, 1467-9868},
  doi = {10.1111/j.1467-9868.2005.00532.x},
  urldate = {2026-05-28},
  copyright = {https://academic.oup.com/journals/pages/open\_access/funder\_policies/chorus/standard\_publication\_model},
  langid = {english}
}

@Manual{Rcore2022,
    title = {R: A Language and Environment for Statistical Computing},
    author = {{R Core Team}},
    organization = {R Foundation for Statistical Computing},
    address = {Vienna, Austria},
    year = {2022},
    url = {https://www.R-project.org/},
  }

@incollection{akaikeInformationTheoryExtension1998,
  title = {Information Theory and an Extension of the Maximum Likelihood Principle},
  booktitle = {Selected Papers of Hirotugu Akaike},
  author = {Akaike, Hirotogu},
  editor = {Parzen, Emanuel and Tanabe, Kunio and Kitagawa, Genshiro},
  year = 1998,
  pages = {199--213},
  publisher = {Springer New York},
  address = {New York, NY},
  doi = {10.1007/978-1-4612-1694-0_15},
  isbn = {978-1-4612-1694-0}
}

@article{hunterVariableSelectionUsing2005,
  title = {Variable Selection Using {{MM}} Algorithms},
  author = {Hunter, David R. and Li, Runze},
  year = 2005,
  month = aug,
  journal = {The Annals of Statistics},
  volume = {33},
  number = {4},
  publisher = {Institute of Mathematical Statistics},
  issn = {0090-5364},
  doi = {10.1214/009053605000000200},
  urldate = {2025-07-12},
  langid = {english}
}

@article{zouAdaptiveLassoIts2006,
  title = {The Adaptive Lasso and Its Oracle Properties},
  author = {Zou, Hui},
  year = 2006,
  month = dec,
  journal = {Journal of the American Statistical Association},
  volume = {101},
  number = {476},
  pages = {1418--1429},
  issn = {0162-1459, 1537-274X},
  doi = {10.1198/016214506000000735},
  urldate = {2026-04-28},
  langid = {english}
}

@article{poignardAsymptoticTheoryAdaptive2020,
  title = {Asymptotic Theory of the Adaptive Sparse Group Lasso},
  author = {Poignard, Benjamin},
  year = 2020,
  month = feb,
  journal = {Annals of the Institute of Statistical Mathematics},
  volume = {72},
  number = {1},
  pages = {297--328},
  issn = {0020-3157, 1572-9052},
  doi = {10.1007/s10463-018-0692-7},
  urldate = {2026-04-28},
  langid = {english}
}

@article{jonesPanTHERIASpecieslevelDatabase2009,
  title = {{{PanTHERIA}}: A Species-level Database of Life History, Ecology, and Geography of Extant and Recently Extinct Mammals: Ecological Archives {{E090-184}}},
  shorttitle = {PanTHERIA},
  author = {Jones, Kate E. and Bielby, Jon and Cardillo, Marcel and Fritz, Susanne A. and O'Dell, Justin and Orme, C. David L. and Safi, Kamran and Sechrest, Wes and Boakes, Elizabeth H. and Carbone, Chris and Connolly, Christina and Cutts, Michael J. and Foster, Janine K. and Grenyer, Richard and Habib, Michael and Plaster, Christopher A. and Price, Samantha A. and Rigby, Elizabeth A. and Rist, Janna and Teacher, Amber and {Bininda-Emonds}, Olaf R. P. and Gittleman, John L. and Mace, Georgina M. and Purvis, Andy},
  editor = {Michener, W. K.},
  year = 2009,
  month = sep,
  journal = {Ecology},
  volume = {90},
  number = {9},
  pages = {2648--2648},
  issn = {0012-9658, 1939-9170},
  doi = {10.1890/08-1494.1},
  urldate = {2026-06-29},
  copyright = {http://onlinelibrary.wiley.com/termsAndConditions\#vor},
  langid = {english}
}

@Book{kleiberAER2008,
  title = {Applied Econometrics with {R}},
  author = {Christian Kleiber and Achim Zeileis},
  year = {2008},
  publisher = {Springer-Verlag},
  address = {New York},
  doi = {10.1007/978-0-387-77318-6},
  url = {https://CRAN.R-project.org/package=AER},
}

\appendix

\section{Derivation of Surrogate Functions}

\subsection{Surrogate for the negative log-likelihood}\label{app:jen_inq}

The negative log-likelihood involves a log of a sum:
\begin{equation*}
    -\ell(\Psi) = -\sum_{i=1}^{n} \log \sum_{g=1}^{G} \pi_g f_g(y_i; \mathbf{x}_i, \psi_g).
\end{equation*}

Define weights $z_{ig}^{(t)}$ as the posterior responsibilities at the current iterate and auxiliary quantities $r_{ig}^{(t)} = \pi_g f_g(y_i; \mathbf{x}_i, \psi_g) / z_{ig}^{(t)}$, so that $\sum_g \pi_g f_g = \sum_g z_{ig}^{(t)} r_{ig}^{(t)}$ with $\sum_g z_{ig}^{(t)} = 1$. Since $-\log(\cdot)$ is convex, Jensen's inequality gives:
\begin{align*}
    -\log \sum_{g=1}^{G} \pi_g f_g(y_i; \mathbf{x}_i, \psi_g)
    &= -\log \sum_{g=1}^{G} z_{ig}^{(t)} r_{ig}^{(t)} \\
    &\leq -\sum_{g=1}^{G} z_{ig}^{(t)} \log r_{ig}^{(t)} \\
    &= -\sum_{g=1}^{G} z_{ig}^{(t)} \left(\log \pi_g + 
       \frac{y_i \theta_{ig} - b(\theta_{ig})}{a(\phi)}\right) + C_\ell^{(t)},
\end{align*}

where $C_\ell^{(t)}$ absorbs all terms constant with respect to $\Psi$. Summing over $i$ yields the surrogate $h_\ell(\Psi|\Psi^{(t)})$ stated in Section~\ref{sec:methods}.

\subsection{Surrogate for the \texorpdfstring{$\pi$}{pi}-SGL penalty}\label{app:dhi}

The penalty $J(\boldsymbol{\beta})$ consists of two non-smooth terms. Each is majorized by applying a first-order Taylor expansion of the concave square root function $\sqrt{u} \leq \sqrt{u^{(t)}} + (u - u^{(t)})/(2\sqrt{u^{(t)}})$.

Setting $u = \sum_{g=1}^G \pi_g^2 \beta_{jg}^2$:
\begin{equation*}
    (1-\alpha)\sqrt{G}\sqrt{\sum_{g=1}^{G} \pi_g^2 \beta_{jg}^2}
    \leq \frac{(1-\alpha)\sqrt{G}}{2} \sum_{g=1}^{G}
    \frac{\pi_g^2 \beta_{jg}^2}{\sqrt{\sum_{g'=1}^{G} \pi_{g'}^{2(t)} \beta_{jg'}^{2(t)}}} + C_1^{(t)}.
\end{equation*}

Setting $u = \pi_g^2 \beta_{jg}^2$:
\begin{equation*}
    \alpha \pi_g |\beta_{jg}| = \alpha\sqrt{\pi_g^2 \beta_{jg}^2}
    \leq \frac{\alpha}{2} \frac{\pi_g^2 \beta_{jg}^2}{\sqrt{\pi_g^{2(t)} \beta_{jg}^{2(t)}}} + C_2^{(t)}.
\end{equation*}

Summing over $j$ and $g$ and combining both terms yields:
\begin{equation*}
    \lambda J(\boldsymbol{\beta}) \leq \lambda \sum_{j=1}^{p}\sum_{g=1}^{G} v_{jg}^{(t)} \pi_g^2 \beta_{jg}^2 + C^{(t)},
\end{equation*}

where $v_{jg}^{(t)}$ is as defined in Equation~\ref{eqn:v_fxn}, and $C^{(t)} = C_1^{(t)} + C_2^{(t)}$.

\section{IRLS update}\label{app:irls}

The combined surrogate $h(\Psi|\Psi^{(t)}) = h_\ell(\Psi|\Psi^{(t)}) + h_J(\boldsymbol{\beta}|\boldsymbol{\beta}^{(t)})$ separates across components, so $\boldsymbol{\beta}_g$ for each $g$ can be updated independently. Taking the derivative with respect to $\boldsymbol{\beta}_g$ and setting it to zero:
\begin{equation*}
    -\frac{1}{a(\phi)}\sum_{i=1}^{n} z_{ig}^{(t)} \mathbf{x}_i \left(y_i - \mu_{ig}\right) 
    + 2\lambda \mathbf{V}_g^{(t)} \pi_g^2 \boldsymbol{\beta}_g = 0,
\end{equation*}

where $\mu_{ig} = b'(\mathbf{x}_i^\top \boldsymbol{\beta}_g)$ is nonlinear in $\boldsymbol{\beta}_g$, 
preventing a closed-form solution. A second majorization is applied via a first-order 
Taylor expansion of $\mu_{ig}$ around $\boldsymbol{\beta}_g^{(t)}$:
\begin{equation*}
    \mu_{ig} \approx \mu_{ig}^{(t)} + \mathsf{v}(\mu_{ig}^{(t)}) \cdot \mathbf{x}_i^\top 
    (\boldsymbol{\beta}_g - \boldsymbol{\beta}_g^{(t)}),
\end{equation*}

where $\mathsf{v}(\mu_{ig}^{(t)}) = b''(\mathbf{x}_i^\top \boldsymbol{\beta}_g^{(t)})$ is the variance 
function evaluated at the current mean. Substituting into the first-order condition:
\begin{equation*}
    \frac{1}{a(\phi)}\sum_{i=1}^{n} z_{ig}^{(t)} \mathsf{v}(\mu_{ig}^{(t)}) \mathbf{x}_i \mathbf{x}_i^\top 
    \boldsymbol{\beta}_g + 2\lambda \mathbf{V}_g^{(t)} \pi_g^2 \boldsymbol{\beta}_g
    = \frac{1}{a(\phi)}\sum_{i=1}^{n} z_{ig}^{(t)} \mathsf{v}(\mu_{ig}^{(t)}) \mathbf{x}_i \tilde{y}_{ig}^{(t)},
\end{equation*}

where the working response is defined as:
\begin{equation*}
    \tilde{y}_{ig}^{(t)} = \mathbf{x}_i^\top \boldsymbol{\beta}_g^{(t)} + 
    \frac{y_i - \mu_{ig}^{(t)}}{\mathsf{v}(\mu_{ig}^{(t)})}.
\end{equation*}

Defining working weights $w_{ig}^{(t)} = z_{ig}^{(t)} \mathsf{v}(\mu_{ig}^{(t)})$ and writing 
$\mathbf{W}_g^{(t)} = \mathrm{diag}(w_{1g}^{(t)}, \ldots, w_{ng}^{(t)})$, this becomes:
\begin{equation*}
    \frac{1}{a(\phi)}\mathbf{X}^\top \mathbf{W}_g^{(t)} \mathbf{X} \boldsymbol{\beta}_g 
    + 2\lambda \mathbf{V}_g^{(t)} \pi_g^2 \boldsymbol{\beta}_g 
    = \frac{1}{a(\phi)}\mathbf{X}^\top \mathbf{W}_g^{(t)} \tilde{\mathbf{y}}_g^{(t)}.
\end{equation*}

Multiplying through by $a(\phi)$ and rearranging gives the closed-form update:
\begin{equation*}
    \boldsymbol{\beta}_g^{(t+1)} = \left(\mathbf{X}^\top \mathbf{W}_g^{(t)} \mathbf{X} 
    + 2\lambda a(\phi) \mathbf{V}_g^{(t)} \pi_g^2\right)^{-1} 
    \mathbf{X}^\top \mathbf{W}_g^{(t)} \tilde{\mathbf{y}}_g^{(t)}.
\end{equation*}

This is a penalized IRLS update, where the penalty matrix $\lambda a(\phi) \mathbf{V}_g^{(t)} \pi_g^2$ 
acts as a component- and covariate-specific ridge-like regularizer. The intercept 
is left unpenalized by setting the first diagonal entry of $\mathbf{V}_g^{(t)}$ to zero.

\section{Derivation of \texorpdfstring{$\lambda_{\max}$}{lambda-max}}\label{app:lambda}

We derive the smallest value of $\lambda$ that sets all non-intercept coefficients to zero at initialization, adapting the approach of \citet{friedmanRegularizationPathsGeneralized2010} to the finite mixture GLM setting with $\pi$-SGL penalty. In the Elastic-Net (EN) setting of \citet{friedmanRegularizationPathsGeneralized2010}, the penalty takes the form:
\begin{equation}
    J_{EN}(\alpha_{EN}, \boldsymbol{\beta}) = \sum_{j=1}^{p} \left[\frac{1}{2}(1-\alpha_{EN})
    \beta_j^2 + \alpha_{EN}|\beta_j|\right]
\end{equation}

Since the $\pi$-SGL surrogate is a weighted ridge (no $\ell_1$ term), this corresponds to the EN penalty with $\alpha_{EN} = 0$. For each component $g$ and covariate $j$, the surrogate weighted ridge penalty maps to the EN penalty with coefficient-specific weight $\pi_g^2V_{jg}^{(t)}$.

Differentiating the surrogate objective Eqn~\ref{eqn:surr_pen} with respect to $\beta_{jg}$ and evaluating at $\beta_{jg} = 0$:
\begin{align*}
    \frac{\partial h}{\partial \beta_{jg}}\bigg|_{\beta_{jg}=0} &=
    -\sum_{i=1}^{n} z_{ig}^{(t)} \cdot \frac{y_i x_{ij} - x_{ij} \mu_{ig}^{(t)}}{a(\phi)}
    + 2\lambda (1-\alpha_{EN}) V_{jg}^{(t)} \pi_g^{2} \beta_{jg} \\
    &+ \lambda \alpha_{EN} V_{jg}^{(t)}\pi_g^2 \text{sign}(\beta_{jg}) = 0 \\
    \implies &-\sum_{i=1}^{n} z_{ig}^{(t)} \cdot \frac{y_i x_{ij} - x_{ij} \mu_{ig}^{(t)}}{a(\phi)}
    + \lambda \alpha_{EN} V_{jg}^{(t)}\pi_g^2 \text{sign}(\boldsymbol{0}) = 0.
\end{align*}

The subgradient condition for $\hat{\beta}_{jg} = 0$ requires:
\begin{equation*}
    \left| \mathbf{X}_j^\top \mathbf{Z}_g^{(0)} \left(\mathbf{y} - 
    \boldsymbol{\mu}_g^{(0)}\right) \right| \leq 
    \lambda \cdot \alpha_{EN} V_{jg}^{(0)} \pi_g^{2(0)},
\end{equation*}

where $\mathbf{X}_j$ is the $j$-th column of $\mathbf{X}$, $\mathbf{Z}_g^{(0)} = \mathrm{diag}(z_{ig}^{(0)})$ contains the initial posterior responsibilities, and $\boldsymbol{\mu}_g^{(0)}$ contains the initial component means. However, $\alpha_\text{EN} = 0$ makes $\lambda_{\max}$ unbounded, so we set $\alpha_\text{EN} = \varepsilon_0$, for a small $\varepsilon_0 > 0$ to ensure numerical stability. 

Similarly, since $\beta_{jg}^{(0)} = 0$ makes $V_{jg}^{(0)}$ undefined, we initialize $\beta_{jg}^{(0)} = \varepsilon_0$. Substituting into $V_{jg}$ and simplifying, the condition for $\hat{\beta}_{jg} = 0$ for all $j, g$ becomes:
\begin{equation*}
    \lambda \geq \max_{j,g} \frac{\left|\mathbf{X}_j^\top \mathbf{Z}_g^{(0)}
    \left(\mathbf{y} - \boldsymbol{\mu}_g^{(0)}\right)\right|}
    {\dfrac{(1-\alpha)\pi_g^{(0)2}\sqrt{G}}{\sqrt{\sum_{g'=1}^{G}
    (\pi_{g'}^{(0)})^2}} + \alpha\pi_g^{(0)}}.
\end{equation*}

To obtain a single $\lambda_{\max}$ valid across all $\alpha \in [0,1]$, note that $D_g(\alpha)$ is linear in $\alpha$ and therefore minimized at an endpoint. For the smallest component the group-lasso endpoint ($\alpha = 0$) is binding, since $\sqrt{G}\pi_{\min}^{(0)} \leq \sqrt{\sum_{g'}(\pi_{g'}^{(0)})^2}$ (as $\pi_{g'}^{(0)} \geq \pi_{\min}^{(0)}$ for all $g'$) gives $D_g(0) = \frac{(\pi_{\min}^{(0)})^2\sqrt{G}}{\sqrt{\sum_{g'}(\pi_{g'}^{(0)})^2}} \leq \pi_{\min}^{(0)} = D_g(1)$. Replacing each denominator by this minimum yields:
\begin{equation*}
    \lambda_{\max} = \frac{\sqrt{\sum_{g'=1}^{G}(\pi_{g'}^{(0)})^2}}
    {\sqrt{G}\,(\pi_{\min}^{(0)})^2} \max_{j,g} 
    \left|\mathbf{X}_j^\top \mathbf{Z}_g^{(0)}\left(\mathbf{y} - 
    \boldsymbol{\mu}_g^{(0)}\right)\right|,
\end{equation*}

where $\mathbf{Z}_g^{(0)}$ and $\pi_g^{(0)}$ are taken from the unregularized model fit. Under equal proportions the leading factor reduces to $1/\pi_{\min}^{(0)}$; it is larger, and hence more conservative. This value is independent of $\alpha$, so a single $\lambda_{\max}$ suffices for the entire grid search over $\alpha \in [0,1]$.

\phantomsection\label{supplementary-material}
\bigskip





\end{document}